%% file: root.tex
\documentclass[letterpaper, 10 pt , conference]{ieeeconf}

\IEEEoverridecommandlockouts                              % This command is only needed if 
\usepackage{graphicx} % for pdf, bitmapped graphics files
\usepackage{epsfig}
\usepackage{epstopdf}
\usepackage{stmaryrd}
\usepackage{cite} % allows to put citations like [1]-[5]
\usepackage{amsmath} % assumes amsmath package installed
\usepackage{amssymb}  % assumes amsmath package installed
\usepackage{mathrsfs}
\usepackage{color}
\usepackage{nicefrac}
\usepackage[linesnumbered, ruled, vlined]{algorithm2e}
\usepackage{subcaption} % for subfigures
\usepackage{mathtools}
\usepackage{bm} % bold symbols
\mathtoolsset{showonlyrefs}
\usepackage[dvipsnames]{xcolor}

\usepackage{pgfplots}
\DeclareUnicodeCharacter{2212}{−}
\usepgfplotslibrary{groupplots, dateplot}
\usetikzlibrary{patterns, shapes, automata, arrows, shapes.arrows, positioning, decorations.pathreplacing, calligraphy, calc}

\pgfplotsset{compat=newest}

\usepackage{tikz}
\usepackage{amsthm}

\usepackage{wrapfig,graphicx}
\usepackage{hyperref}
\graphicspath{{.}}

\theoremstyle{plain}
\newtheorem{thm}{Theorem}

\newtheorem{lemma}{Lemma}
\newtheorem{problem}{Problem}
\newtheorem{remark}{Remark}
\newtheorem{assumption}{Assumption}
\newtheorem{mechanism}{Mechanism}
\usepackage{etoolbox}
\usepackage{float}
\usepackage{tabularx}

\def\th@definition{
	\thm@headfont{\itshape} % Heading font is italic
	\thm@notefont{} % Note is same as heading
}
\makeatother
\usepackage{color}

\theoremstyle{definition}
\newtheorem{definition}{Definition}

\newcommand{\diag}{\textnormal{diag}}

\makeatletter
\patchcmd{\@makecaption}
{\scshape}
{}
{}
{}
\makeatother

\makeatletter
\def\endthebibliography{%
	\def\@noitemerr{\@latex@warning{Empty `thebibliography' environment}}%
	\endlist
}
\makeatother

\title{\LARGE \bf
Differentially Private Computation of Basic Reproduction Numbers \\
in Networked Epidemic Models
}

\author{Bo Chen$^{\ast}$, Baike She$^{\ast}$, Calvin Hawkins$^{\ast}$, Alex Benvenuti$^{\ast}$, Brandon Fallin$^{\ast}$, Philip E. Par\'e$^{\dagger}$, Matthew Hale$^{\ast}$% <-this % stops a space
\thanks{$^{\ast}$Department of  Mechanical and Aerospace Engineering at the University of Florida, Gainesville, FL USA. Emails: \texttt{\{bo.chen,shebaike,calvin.hawkins,abenvenuti,brandon} \texttt{fallin,matthewhale\}@ufl.edu}. This work was supported by NSF CAREER grant 1943275, AFRL grant FA8651-23-F-A008, AFOSR grant FA9550-19-1-0169, ONR grant N00014-21-1-2502, and DARPA grant HR00112220038.
}
\thanks{$^{\dagger}$Elmore Family School of Electrical and Computer Engineering at Purdue University, West Lafayette, IN USA. Email: \texttt{philpare@purdue.edu}. This work was supported by NSF CAREER ECCS 2238388.
}
}

\begin{document}

\maketitle

\input{abstract}
%input{keywords}
\input{intro}

\input{Background}

%\input{privacy_mechanism_v1}
%\input{privacy_mechanism_v2}
\input{Lvl_of_Pen}

\input{privacy_mechanism_v3}
\input{CaseStudy_real_data}
\input{Conclusions}

%%%%%%%%%%%%%%%%%%%%%%%%%%%%%%%%%%%%%%%%%%%%%%%%%%%%%%%%%%%%%%%%%%%%%%%%%%%%%%%%

\bibliographystyle{IEEEtran}{}
\bibliography{root}
%\newpage
%\input{Appendix}
\input{Appendix_new}
%\newpage
%\red{
%\begin{enumerate}
%\item Verify that all inequality arguments are valid. For example, suppose~$x_1 \leq y \leq x_2$, and suppose we want to derive a sufficient
%condition to have~$y \leq z$. It is sufficient to have~$x_2 \leq z$, but it is not sufficient to have~$x_1 \leq z$. 
%\item Make sure every symbol is defined, ideally before it is used. Pay special attention to network-level terms (like~$H$,~$V$, etc.). 
%\item Look at all margin comments and make the appropriate changes.
%\item Only use~$\tilde{y}$ to denote the privatized output. Specifically, only write~$\Delta_{\ell_2} y$ and don't write~$\Delta_{\ell_2}\tilde{y}$
%\item Put all assumptions in the theorem statement, not in the proof
%\end{enumerate}
%}

\end{document}

%% file: abstract.tex
\begin{abstract}
    The basic reproduction number of a 
    networked epidemic model, denoted $R_0$,
    %i.e., the maximal absolute eigenvalue of a graph's weight matrix, 
    can be computed from a network's topology to quantify epidemic
    spread. 
    %plays an important role, serving to both predict the spread of an epidemic and
    %as a means
    %to convey valuable pandemic information to the general population. 
    However, disclosure of $R_0$ risks revealing 
    sensitive information about the underlying network, such as 
    an individual's relationships 
    within a social network. 
    %individuals, such as past travel histories.
    Therefore, we propose a framework to compute and release $R_0$ 
    in a differentially private way. 
    %The input perturbation mechanism adds carefully calibrated noise to the matrix entries before computing $R_0$, whereas the output perturbation mechanism directly adds noise to the value of $R_0$. For both mechanisms, 
    First, we provide a new result that shows how~$R_0$ can be used to bound
    the level of penetration of an epidemic within a single community as a motivation for the need of privacy, which
    may also be of independent interest. We next develop a privacy mechanism to formally safeguard the 
    edge weights in the 
    underlying network when computing~$R_0$.
    Then we formalize tradeoffs between the level of privacy and the accuracy of 
    values of the 
    privatized $R_0$. 
    To show the utility of the private~$R_0$ in practice, we use it
    to bound this level of penetration under privacy, 
        and concentration bounds on these analyses show
    they remain accurate with privacy implemented. We apply our results to
    real travel data gathered during the spread of COVID-19, and we show that, under real-world conditions,
    we can compute~$R_0$ in a differentially private way while incurring errors as low as~$7.6\%$ on average. 
    %To provide a practical illustration, 
    %we apply privacy to $R_0$ and evaluate the resulting accuracy loss, particularly in the context of the level of penetration within one community, with concentration bounds. Simulation results confirm the utility of private $R_0$ in these contexts.
\end{abstract}

%% file: intro.tex
\vspace{-2ex}
\section{Introduction}
\vspace{-1ex}
Compartmental epidemic models have been used to model the spread of epidemics, assess pandemic severity, predict spreading trends, and facilitate policy-making~\cite{brauer2008compartmental}. This progress has been in part propelled by advancements in network science
%improvements in computational hardware capabilities, and the diversification of methods for collecting and storing detailed data on disease transmission
\cite{pare2020modeling, she2021peak, mei2017dynamics, brockmann2013hidden}. 
%In particular, networked spreading models have played a crucial role in addressing the global spread of epidemics~\cite{brockmann2013hidden}. However, 
%unlike researchers who have access to more comprehensive information about epidemic spread, including data on disease transmission, population movement, and hospital records, 
Due to their complexity, it can be difficult to communicate the 
intricate details and conclusions of these models~\cite{bengio2020need},
though the basic reproduction number of a spreading process has emerged as 
one concise way to convey information about the spread of epidemics~\cite{delamater2019complexity,aronson2020will}. 
%To bridge this gap and convey valuable pandemic information to the general public, as well as to garner support for intervention strategies, there is a need for a powerful yet concise means of describing the process of epidemic transmission.

The basic reproduction number of a spreading process, denoted~$R_0$, is the average number of individuals that an infected person will infect in a fully susceptible population~\cite{delamater2019complexity}. 
Intuitively, higher $R_0$ values indicate greater transmissibility.
For example, the basic reproduction numbers for diseases like measles, SARS-CoV-1, and the Ebola virus are approximately 14.7, 3.1, and 1.9, respectively~\cite{aronson2020will}. 

Researchers have 
defined basic reproduction numbers for networked epidemic models~\cite{pare2020modeling}, which capture not only the transmissibility of the epidemic process but also the effect of the graph structure. For example, in a networked susceptible-infected-susceptible (SIS) model, basic reproduction numbers less than or equal to~$1$ ensure that the size of the infected population eventually converges to zero~\cite{pare2020modeling}. 
Thus,~$R_0$ can be used to forecast the future behavior
of an epidemic and communicate with the public in a concise way. 
%\textcolor{red}{In this paper, we present a new way to use~$R_0$ to bound
%the level of penetration of an epidemic into a community, which
%may be of independent interest. %Specifically, we 
%bound the size of the uninfected population in a community
%at equilibrium, and this bound only requires~$R_0$ to compute.}

Unfortunately, it is well-known that sharing even scalar-valued graph properties like~$R_0$ can pose privacy threats~\cite{imola2021locally,Karwa2014Private,Day2016Publishing,ZHANG2021Differentially,chen2021edge}. 
In particular, one can initiate a \emph{reconstruction attack}, in which an attacker combines released graph properties (here, $R_0$) with other information to reconstruct the underlying graph information, such as 
%the absence or presence of a node or a edge, as well as 
the weights in a weighted graph, which can be sensitive. 
%The disclosure of such information may in turn reveal sensitive information about individuals in them.
For example, consider a residential community of a small number of households, whose interactions with other communities contribute to the modeling of graph weights. Then one may be able to infer
the travel habits of a person by reconstructing these graph weights; see~\cite{imola2021locally,Karwa2014Private,Day2016Publishing,ZHANG2021Differentially,chen2021edge}
for additional discussion of privacy threats for graphs. 
In addition, this type of privacy risk extends 
to large regions as well~\cite{Bureau2023DP}.
Thus, despite the importance of $R_0$, it is undesirable to
publish $R_0$ without any protections. 

In this work, we provide these protections by using differential 
privacy~\cite{Sealfon2016shortest,pinot2018clustering} to protect 
graph weights when computing $R_0$. %The next generation matrix is \textcolor{red}{some stuff}, and~$R_0$
%\textcolor{red}{depends on it in some way}.
%The value of $R_0$ is closely related to the next generation matrix. 
%is equal to the spectral radius, i.e., the maximal absolute eigenvalue, of a network's weight matrix. 
Our implementation uses an input perturbation approach, which 
%%Specifically, input perturbation 
%provides differential privacy guarantees by 
first adds noise directly to the matrix
of graph weights, then
%directly to the next generation
%matrix, thereby privatizing it, and then 
computes~$R_0$ from this private matrix. 
%%while output perturbation is by directly adding noise into the value of $R_0$. 
%
Differential privacy provides strong, formal privacy protections for sensitive
data, and it is desirable here because differentially private
data may be freely post-processed without harming its guarantees~\cite{dwork2014algorithmic}.
In particular, after privatizing the matrix of weights, we can 
compute~$R_0$ and use it for epidemic forecasting without harming privacy. 

To ensure that private values of $R_0$ 
enable useful analyses, we use 
%bounded mechanisms, namely the bounded Laplace mechanism~\cite{holohan2018bounded} and 
the bounded 
Gaussian mechanism~\cite{chen2022bounded}, which only generates private outputs within specified ranges.
We follow this approach because~$R_0$ and graph weights are non-negative, which ensure that their private
forms are as well. 
% 
%Both mechanisms ensure the private values are informative for the intended analysis by only generating private outputs lie within specific ranges.
Moreover, as a motivating example, we present a new way to use~$R_0$ to bound
the level of penetration of an epidemic into a community, which
may also be of independent interest. Specifically, we 
bound the size of the uninfected population in a community
at equilibrium, and this bound is a function of only~$R_0$.

Our specific contributions in this work are: 
\begin{enumerate}
    \item A result to use values of~$R_0$ to analyze
    the spread of an epidemic in terms of the eventually remaining susceptible population.
    \item A mechanism for differential privacy that protects the underlying graph weights when publishing the basic reproduction number $R_0$.
    \item Privacy-accuracy tradeoffs that quantify both  
    (i) the expected deviation from the true value of $R_0$ and 
    (ii) the accuracy of predictions of the remaining susceptible population as functions of the strength of privacy.   
    %\item A framework to use private values of~$R_0$ to analyze
   %the bounds on the error that privacy introduces into this the analysis of the eventually remaining susceptible population of an epidemic.
\end{enumerate}

%Numerical experiments demonstrate that the private~$R_0$ maintains accuracy for useful analysis even with strong privacy implemented.
We use travel data from Minnesota during the COVID-19 pandemic 
show that a real-world deployment
of this privacy framework leads to errors as low as~$7.6\%$ on average.

\textbf{Relation to prior work: } There exist numerous differential privacy implementations for graph properties, including counts of sub-graphs and triangles~\cite{imola2021locally,Karwa2014Private}, degree distributions~\cite{Day2016Publishing,ZHANG2021Differentially}, and algebraic connectivity~\cite{chen2021edge,hawkins2022node}. 
In many of these prior works, differential privacy has been applied with edge and node adjacency~\cite{Hay2009accurate,blocki2013differentially,Kasiviswanathan2013Analyzing} to obfuscate the absence and/or
presence of a pre-specified number of edges or nodes. 
In contrast, we consider
graphs with node and edge sets that are publicly known. 
We do so 
because networked epidemic models often
use vertices to represent communities and/or cities and use edges 
to represent
connections such as highways or flights,
all of which are publicly known. 
We instead use weight adjacency~\cite{Sealfon2016shortest}
and protect the weights in a weighted graph. 

Differential privacy has been used to protect the eigenvalues of certain types of matrices~\cite{Wang2013Differential,chen2021edge,hawkins2022node}. 
We differ by privatizing matrices of weights in weighted graphs, which
those works do not consider. 
%implements output perturbation of differential privacy with post-processing in the eigendecomposition of a graph’s adjacency matrix. There is also work~\cite{chen2021edge,hawkins2022node} on the output perturbation for the second smallest eigenvalue of a graph's Laplacian matrix. By contrast, our implementations focus on weight matrices that are symmetric with non-negative entries and our output perturbation mechanism does not require post-processing. More closely related to input perturbation, 
Work in~\cite{Chanyaswad2018MVG} adds
noise drawn from a matrix-variate Gaussian distribution to a matrix for privacy protection. However, such noise is unbounded and our work instead adds bounded noise to ensure 
that privatized weights and values of~$R_0$ remain non-negative.

%% file: Background.tex
\vspace{-1ex}
\section{Background and Problem Formulation}\label{sec:problemFormulation}
\vspace{-1ex}
In this section, we introduce notation, background on epidemic models and privacy, and problem statements. 
\vspace{-2ex}
\subsection{Notation}
\vspace{-1ex}
We use $\mathbb{R}$ to denote the real numbers, $\mathbb{R}_{\geq 0}$ to denote the non-negative reals, and $\mathbb{R}_{> 0}$ denote the positive reals. 
For a random variable $X$, $\mathbb{E}[X]$ denotes its expectation and $\text{Var}[X]$ denotes its variance. Let $\mathbf{1}_{T}(\cdot)$ denote the indicator function of set $T$.
%Let $\text{Uniform}(a,b)$ denote the uniform distribution between $[a,b]$ with $-\infty<a<b<\infty$. 
We use $[n]$ 
%\bo{Baike, in our previous paper we usually use $[n]$ to denote this set. It is not a big deal though lol just curious is it universal to use $[n]$ in epidemic research?} 
to denote $\{1,2, \dots, n\}$.
For a real square matrix $M$, we use $\rho(M)$ to denote
its spectral radius. For any two matrices $A,B\in\mathbb{R}^{n\times n}$,
we write $A\geq B$ if $a_{ij}\geq b_{ij}$, %$v>w$ if $v_{i}>w_{i}$,  
$A> B$ if $a_{ij}\geq b_{ij}$ and $A\neq B$, 
and $A\gg B$ if $a_{ij}>b_{ij}$, for all $i,j\in [n]$.
%\bo{Baike, please complete this sentence.}
These comparison notations between matrices apply to vectors as well.
For a vector~$v \in \mathbb{R}^n$, we write~$\textnormal{diag}(v)$
to denote the diagonal matrix whose~$i^{th}$ diagonal
entry is~$v_i$ for each~$i \in [n]$. We use $||\cdot||_F$ to denote the Frobenius norm of a matrix.

Let $[a,b]^n$ be the Cartesian product of $n$ copies of the same interval $[a,b]$. For graphs, let $G=(V,E,W)$ denote an undirected, connected, and weighted graph with node set $V$, edge set $E$, and weight matrix $W$, 
where $w_{ij} \geq 0$ denotes the $i^{th},j^{th}$ entry of the weight matrix $W$. Let $|\cdot|$ denote the cardinality of a set. 
For a given weight matrix~$W$, 
we use $n_w=|\{w_{ij}>0 : i, j \in [n]\}|$ 
to denote the number of positive entries in $W$. We use $\mathcal{G}_n$ to denote a set of all possible undirected, connected, weighted graphs $G$ on $n$ nodes. We also use the special functions
\begin{align}
    \varphi(x) &= \frac{1}{\sqrt{2\pi}}\exp\left(-\frac{1}{2}x^2\right),\label{eq:gaussian_pdf}\\
    \Phi(x) &= \frac{1}{2}\left(1+\frac{2}{\sqrt{\pi}}\int_0^{\frac{x}{\sqrt{2}}} \exp(-t^2)dt\right)\label{eq:gaussian_cdf},
    %\text{erf}(x) &:= \frac{2}{\sqrt{\pi}}\int_0^x \exp(-t^2)dt,\label{eq:error_function}
\end{align}
which are the probability density function and the cumulative distribution function of the standard normal distribution, respectively.
%and~\eqref{eq:error_function} denotes the error function, .
\vspace{-2ex}
\subsection{Networked Epidemic 
\vspace{-1ex}Models}\label{sec:prelim_epidemic}
We consider networked susceptible-infected-susceptible (SIS) and susceptible-infected-recovered (SIR) models.
%SIS \bo{or SIR} model. 
Let $\Bar{G}=(V,E,B)\in\mathcal{G}_n$ denote a 
connected and undirected spreading network that models an epidemic spreading process over $n$ connected communities. 
Let $V$ and $E$ denote the communities and the transmission channels between these communities, respectively.
We use $s(t), x(t), r(t)\in[0,1]^n$ to represent the susceptible, infected, and recovered state vectors, respectively.
That is, for all~$i \in [n]$, the value of~$s_i(t) \in [0, 1]$ is the portion of the population of community~$i$ that is susceptible at time~$t$;
the values of~$x_i(t)$ and~$r_i(t)$ are the sizes of the infected and recovered portions of community~$i$, respectively. 
We use $B\in \mathbb{R}^{n\times n}_{\geq 0}$, with $b_{ij}\in[0,1]$ for all~$i, j \in [n]$, to denote the transmission matrix and $\Gamma=\text{diag}(\gamma_1,\gamma_2,\dots,\gamma_n)$, with $\gamma_i>0$ for all~$i \in [n]$, to denote the recovery matrix. Thus, 
the value of $b_{ij}$ captures the transmission process from the community $j$ to community $i$, while $\gamma_i$ captures the recovery rate of community $i$. The  networked SIS and  SIR models are 
\begin{equation}
    \begin{cases}\label{eq:SIS}
       \dot{s}(t) &= -\diag(s(t))Bx(t) + \Gamma x(t), \\
        \dot{x}(t)  &= \diag(s(t))Bx(t)-\Gamma x(t), 
    \end{cases}
\end{equation}
and 
\vspace{-2ex}
\begin{equation}
    \begin{cases}\label{eq:SIR}
     \dot{s}(t)  &= -\diag(s(t))Bx(t), \\
     %\label{eq:SIR_S}\\
        \dot{x}(t)  &= \diag(s(t))Bx(t) - \Gamma x(t),\\
        %\label{eq:SIR_I}\\
        \dot{r}(t)  &= \Gamma x(t),
        %\label{eq:SIR_R},
    \end{cases}
\end{equation}
respectively. For all $i\in[n]$,
%$s_i(t), x_i(t), r_i(t)\in[0,1]$, $i\in[n]$, and 
$s_i(t)+x_i(t)+r_i(t)=1$~\cite{pare2020modeling}.
%In addition, networked SIS and SIR models admit the same dynamics for infected states.

%The next generation matrix was developed by epidemiologists to formalize the basic reproduction numbers of epidemic compartmental models~\cite{diekmann2010construction}. 
For networked SIS and SIR spreading models, researchers have defined 
\textit{the next generation matrix} $W = \Gamma^{-1}B$
% \begin{equation*}
%     R_0 = \rho(W) %\label{eq:def_r_0},
% \end{equation*}
to characterize the global behavior of networked SIS and SIR models in~\eqref{eq:SIS} and~\eqref{eq:SIR}~\cite{pare2020modeling,mei2017dynamics,she2021peak}. 
One can then compute the basic reproduction number from~$W$ via $R_0 = \rho(W)$. %where~$\rho(\cdot)$ denotes the spectral radius. 
%For instance, if $R_0 = \rho(W)$ of a networked SIS spreading model is smaller than or equal to 1, the infected states $x(t)$ will converge to zero eventually, and none of the community will have infected individuals. 

\begin{remark} \label{rem:pen}
Developments in~\cite{mei2017dynamics,zhen2023steady} 
suggest that the basic reproduction number in compartmental models is linked to the remaining susceptible population at the disease-free equilibrium, which represents the \textit{level of penetration} in a community. This level of penetration quantifies the virus' impact, namely
how many individuals will become infected. %As a scalar value, the reproduction number provides substantial insights for researchers studying epidemic spread across connected communities, both collectively and individually. 
\end{remark}

To safeguard the weights in~$W$, it is essential to provide privacy for~$W$ when publishing $\rho(W)$. 
% indicate that the basic reproduction number of compartmental models is related to the remaining susceptible population at the disease-free equilibrium,  where the remaining susceptible population within the community can be defined as~\textit{the level of penetration} of that  community. 
% The level of penetration captures the damage caused by the virus such that how many people will be infected after the spread. 
% Hence, as a scalar, the reproduction number itself can infer abundant information for researchers to study the epidemic spreading over the connected communities overall and individually. In order to protect the detailed spreading information, it is necessary to maintain some level of privacy for $\rho(W)$ before publishing.
Since $R_0$ is defined in terms of $W$ rather than $B$, we will privatize $W$ directly. 
To reflect our focus,
we define a weighted graph for a spreading network as $G=(V,E,W)$, with $W=\Gamma^{-1}B$, 
and we focus on this class of graphs going forward.

\subsection{Differential Privacy}
\vspace{-1ex}
Differential privacy is enforced by a randomized map, called a privacy 
\emph{mechanism}, which must
ensure that nearby inputs to the mechanism produce outputs that are statistically approximately indistinguishable from each other. 
%In formal terms, a mechanism must blur distinctions between \emph{adjacent} inputs. While most literature uses edge and node adjacency for graphs, these notions may not suit epidemic control problems well. This is because, in graphs where nodes represent groups of individuals, the existence of a node or edge might be predetermined. However, edge weights can contain sensitive information about an individual, derived from factors like close friends and interaction frequency. 
In this paper, we adopt weight adjacency~\cite{Sealfon2016shortest}, which 
formalizes the notion of ``nearby'' for weighted graphs. 
\begin{definition}\cite{Sealfon2016shortest}\label{def:adjacency}
    Fix an undirected weighted graph $G=(V,E,W)\in\mathcal{G}_n$. Then another undirected weighted graph $G'=(V,E,W')$ is \emph{weight adjacent} to $G$, denoted $G\sim G'$, if 
    %$||W-W'||_F=\sqrt{\sum_{i=1}^{n}\sum_{j=1}^n(w_{ij}-w_{ij}')^2}\leq k$,
    %\begin{equation}
        $||W-W'||_F = \sqrt{\sum_{i=1}^{n}\sum_{j=1}^n(w_{ij}-w_{ij}')^2} \leq k$,
        %||W-W'||_1=\sum_{i,j}|w_{ij}-w_{ij}'|\leq k,
        %\sqrt{\sum_{i=1}^{n}\sum_{j=1}^n(w_{ij}-w_{ij}')^2}\leq k,
    %\end{equation}
    where $k>0$ is a user-specified parameter. \hfill $\lozenge$
\end{definition}
%\baike{[Baike: A question here: Does this definition validate for graphs that have weighted edges, i.e., not just 0s and/or 1s?]} \bo{Yes, it is valid for all graphs with nonnegative weights.}

Definition~\ref{def:adjacency} states that two graphs are weight adjacent if they have the same edge 
and node sets, 
and the distance between their weight matrices is bounded by~$k$ in the Frobenius norm. 
We next introduce the definition of differential privacy in the form in which we will
use it in this paper. 

\begin{definition}[Differential Privacy\cite{dwork2014algorithmic}]\label{def:differential_privacy}
    Let $\epsilon > 0$ be given and fix a probability space $(\Omega, \mathcal{F}, \mathbb{P})$. 
    Then a mechanism $\mathcal {M}: \Omega \times \mathbb{R}^{n \times n}_{\geq 0} \rightarrow \mathbb{R}^{n \times n}_{\geq 0}$ is $\epsilon$-differentially private if,
    for all weight adjacent graphs $G=(V,E,W)$ and $G'=(V,E,W')$ in~$\mathcal{G}_n$, it satisfies $\mathbb{P}\big[\mathcal{M}(W) \in S\big] \leq e^{\epsilon} \cdot \mathbb{P}\big[\mathcal{M}\left(W'\right) \in S\big]$
    %\begin{equation}
        %\mathbb{P}\big[\mathcal{M}(W) \in S\big] \leq \exp (\epsilon) \cdot \mathbb{P}\big[\mathcal{M}\left(W'\right) \in S\big]
    %\end{equation}
    for all sets $S$ in the Borel $\sigma$-algebra over $\mathbb{R}^{n \times n}_{\geq 0}$. \hfill $\lozenge$
\end{definition}

The privacy parameter $\epsilon$ controls the strength of privacy and a smaller $\epsilon$ implies stronger privacy. Differential privacy even with large $\epsilon$, e.g., $\epsilon>10$, provides much stronger empirical privacy than no differential privacy~\cite{jagielski2020auditing,nasr2021adversary,song2019auditing,balle2022reconstructing}. In this paper, we provide differential privacy through input perturbation. 
For a weighted graph $G=(V,E,W)$, input perturbation first privatizes~$W$ itself by randomizing it,
then computes~$R_0$ from the private~$W$. 
Due to differential privacy's immunity to post-processing, the resulting~$R_0$ is also differentially private. 

%\baike{[BS: We need to say somewhere explicitly that we add privacy on the networked captured by $W = \Gamma^{-1}B$. I realize this connection until now. Previously, I thought W is a general notation for weighted network but we use define $W = \Gamma^{-1}B$ to represent the graph we want to study.]} \bo{Could you elaborate on this?}
\vspace{-1ex}
\subsection{Setup for Private Analysis} 
\vspace{-1ex}
\label{ss:setup}
In this subsection, we formalize the information that the sensitive graph $G$ discloses to epidemic analysts and the information it should conceal.
%We then illustrate this setting with an example.

% We assume that epidemic analysts have access to the structure of a given graph. Specifically, for the generalized graph $G=(V,E,W)$ with $W=\Gamma^{-1}B$, we consider the vertices set $V$ and edges set $E$ as publicly available information. 
% %This assumption is often applicable in networked epidemic models, where the vertices represent communities or cities, and the edges correspond to geographical relationships between them, including factors such as locations or highways/flights connecting them, which are typically publicly known. 
% Additionally, we assume the graph curator does not trust the epidemic analysts, hence both the transmission matrix $B$ and the recovery matrix $\Gamma$, as well as the next generation matrix $W$, will not be shared with them.
%\baike{[usually weights of $W$ cannot be obtained directly. In mathematical analysis, we can obtain B and $\Gamma$ separately. We need to formulate the problem as adding noise to $W$ instead of $B$.]}
%The disclosure of these graph weights can potentially increase the risks of individual data leakage, particularly related to sensitive information such as travel details. 
We assume epidemic analysts have access to a graph's vertex set $V$ and edge set $E$. However, we do not share the transmission matrix $B$, the recovery matrix $\Gamma$, or the next generation matrix $W$ with them since these are sensitive. 
In addition, it is well-known that publishing even scalar-valued graph properties can pose substantial privacy threats~\cite{imola2021locally,Karwa2014Private,Day2016Publishing,ZHANG2021Differentially,chen2021edge}.
%For example, one can initiate a \emph{reconstruction attack}, in that an attacker may combine the released graph properties ($R_0$ in our case) with other information to reconstruct the underlying sensitive graph information (graph weights $W$ in our case). 
As a result, the value of~$R_0$ is not shared with epidemic analysts either. 
Instead, they will \emph{only} receive a differentially private version of $R_0$, denoted by $\tilde{R}_0$.

Lastly, we assume that each entry $w_{ij}$ lies in an interval $(\underline{w}_{ij},\Bar{w}_{ij}]$, where $\underline{w}_{ij}$ and $\Bar{w}_{ij}$ are known lower and upper bounds and will be shared with analysts. It should be noted that while sharing these bounds conveys some information about the underlying graph, it is not highly sensitive information. Other publicly available data sources or databases, such as the number of highways connecting communities or community population statistics, can be used to infer 
information of this kind. In practice, one can therefore 
group values of $w_{ij}$ into certain ranges without harming privacy, which is possible precisely
because approximate ranges of these values can be inferred from publicly available data. 
%The value of $\min_i\gamma_i$ is assumed to be publicly available through, for example, a public health authority, since the recovery rate is related to the days of being infectious after infection~\cite{klopfenstein2020features}. 
%By Ger\v{s}gorin's circle theorem \cite[Fact 4.10.16.]{Bernstein2005MatrixMT}, we have $R_0 \in (0, \Bar{R}_0]$, where $\Bar{R}_0=n\Bar{w}$ and $n$ is the number of nodes in the network.

%\baike{[Baike: Is $n$ the number of nodes in the network of interest?]}\bo{Yes}
\vspace{-1ex}
\subsection{Problem Statement}
We next state the problems that we will solve.
\vspace{-1ex}
\begin{problem}\label{prob:level_of_pen}
    Build an upper bound on the \emph{level of penetration} of a community (in the sense of Remark~\ref{rem:pen})
    within a spreading network by using its basic reproduction number $R_0$.
\end{problem}

\begin{problem}\label{prob:input_perturbation_design}
    Develop a differential privacy mechanism to provide differential privacy in the sense of Definition~\ref{def:differential_privacy} for the next generation matrix $W$ 
    when computing~$R_0$.
\end{problem}

% \begin{problem}\label{prob:output_perturbation_design}
%     Develop an \textbf{output perturbation} mechanism to provide differential privacy guarantee in the sense of Definition~\ref{def:differential_privacy} for the next generation matrix $W$ while publishing the reproduction number $R_0$.
% \end{problem}

\begin{problem}\label{prob:accuracy_private_r0}
    Given a reproduction number $R_0$, for private values $\Tilde{R}_0$ generated by the proposed mechanism, develop bounds on the expected accuracy loss $\mathbb{E}[|\tilde{R}_0-R_0|]$ of the developed mechanism as a function of privacy level.
\end{problem}

\begin{problem}
    Analytically evaluate the utility of the private reproduction number $\tilde{R}_0$ in modeling the level of penetration of networked spreading processes.
\end{problem}
\vspace{-2ex}
\subsection{Probability Background}
\vspace{-1ex}
%We give some theorems that will be used in later sections. %We first introduce the definition of the \emph{sub-Gaussian} random variable.

%We introduce the \emph{truncated Gaussian} random variable.
\begin{definition}\cite{burkardt2014truncated}
    The \emph{truncated Gaussian} random variable, written as $\text{TrunG}(\mu,\sigma,l,u)$, that lies within the interval $(l,u]$, where $-\infty< l < u< +\infty$, and centers on $\mu\in(l,u]$ is defined by the probability density function $p_{TG}$ with
    \begin{equation}
        p_{TG}(x) = \begin{cases}
        \frac{1}{\sigma} \frac{\varphi\left(\frac{x-\mu}{\sigma}\right)}{\Phi\left(\frac{u-\mu}{\sigma}\right)-\Phi\left(\frac{l-\mu}{\sigma}\right)} & \text{if } x\in (l,u] \\
        0 & \text{otherwise,}
    \end{cases}
    \end{equation}
    and $\sigma>0$, where $\phi(\cdot)$ is from~\eqref{eq:gaussian_pdf} and $\Phi(\cdot)$ is from~\eqref{eq:gaussian_cdf}.\hfill $\lozenge$
\end{definition}

%For a random variable $Y\sim\text{TrunG}(\mu,\sigma,l,u)$, $\mathbb{E}[Y]=\mu$ if and only if $\mu=\frac{l+u}{2}$. 
%\emph{Proof:} See Appendix~\ref{apdx:proof_trunc_gauss_is_sub_gauss}.\hfill $\blacksquare$

%% file: Lvl_of_Pen.tex
\vspace{-2ex}
\section{Penetration Analysis with $R_0$} 
\vspace{-1ex}
\label{Sec:Lvl_of_pen} 
%One of the motivations for privately computing~$R_0$ is to conceal detailed information about individuals.
In this section, we illustrate the value of~$R_0$ in epidemic analysis
by demonstrating one type of information that can be obtained from~$R_0$.
%the impact of the basic reproduction number $R_0$ on the level of penetration into a community within the epidemic spreading network.
As previously mentioned in the problem formulation, it is possible to use~$R_0$ to infer
%estimate 
the remaining susceptible population within a community, referred to as the \emph{level of penetration} of an epidemic. 
This information enables us to determine the total
number of individuals within a given community
who will be infected by a virus over time. 
%\mh{How much of this is new? Be very specific about how relate to [25].}
%To illustrate the effectiveness of the privatized reproduction number, we establish a correlation between the level of penetration into a community within the network and the reproduction number.
%\subsection{Level of Penetration}

% In this subsection, we discuss the impact of the  reproduction number $R_0$ 
% on the level of penetration into a community in the epidemic spreading network. 
% As we discussed in the introduction, it is possible infer the remaining susceptible population within a community through the reproduction number, such that we can obtain that at least how many people will be infected by the virus within one community.
%In particular, we bridge the gap between the basic reproduction number $R_0$ and the existing susceptible population within a community at a disease-free equilibrium %\colorbox{BurntOrange}{[define disease-free equilibrium]}.
% In order to show the efficiency of the private reproduction number, we further build the connection between the level of penetration into a community within the network and the reproduction number in this subsection.  
%the connection developed in this subsection will lay of foundation for leveraging private $R_0$ to protect the level of penetration within a community in the network.
%Then, we explore how to leverage the private R0 to project the level of penetration. Then, we explore the how to leverage the private $R_0$ to project the level of penetration. 

In particular, we will 
quantify the relationship between $R_0$ and the proportion of the susceptibles within community $i$ at a disease-free equilibrium, denoted $s^*_i$, for all $i\in [n]$\footnote{Note that a simulation of~\cite{zhen2023steady} studies the susceptible proportion within a community $i$, $i\in [n]$, at the disease-free equilibrium through a different way of defining the reproduction number of a networked spreading process, i.e., $R_0 = \rho(B\Gamma^{-1})$. In addition,~\cite{zhen2023steady} applies its developed results to networked epidemic spreading dynamics without proving that the networked spreading models satisfy the conditions on its developed results.}.
To do so, we first rewrite the dynamics of %the infected states in 
the networked SIS and SIR models in~\eqref{eq:SIS} and~\eqref{eq:SIR} each with two separate components: 
(i) nonlinear dynamics~\cite[Eq.(2)]{zhen2023steady} to model the susceptible states $s(t)$, which are
%\mh{Do we need to introduce~$v$ and~$u$ if they are just~$x$ and~$z$? Also, can we spell out what~$f$ is? Currently it's just a symbol.}
\vspace{-1ex}
\begin{align}
\label{eq:non_dyn_epi_s}
    \dot s(t) &= f(s(t),x(t)), \\
    u(t) &= I\diag\{s(t)\}Bx(t);
\vspace{-2ex}
\end{align}
(ii) linear dynamics~\cite[Eq.(3)]{zhen2023steady} with external input to model the infected states $x(t)$, which are
\vspace{-1ex}
\begin{align}
\label{eq:non_dyn_epi_x}
    \dot x(t)&=-\Gamma  x(t) + Iu(t),\\  
    y(t) &= Ix(t).
\vspace{-3ex}
\end{align}
% Further, we have the coupled dynamics through~\cite[Eq.(4)]{zhen2023steady}:
% \begin{align}
% \label{eq:non_dyn_epi}
%     v(t)&=y(t)=x(t),\\
% u(t)&=z(t)=\diag\{s(t)\}Bx(t).
% \end{align}
where $I$ is the identity matrix.
We use the coupled dynamics in~\eqref{eq:non_dyn_epi_s}-\eqref{eq:non_dyn_epi_x} to capture the networked $SIS$ 
models, where $f(s(t),x(t)) = -I\diag{s(t)}Bx(t)+\Gamma x(t)$. Similarly, when $f(s(t),x(t)) = -I\diag{s(t)}Bx(t)$, we use ~\eqref{eq:non_dyn_epi_s}-\eqref{eq:non_dyn_epi_x} to represent
SIR models, where  $r(t)=1-s(t)-x(t)$ is omitted.   %Similarly, %when $v(t)$ represents $x(t)$, and $r(t)=1-s(t)-x(t)$,

\vspace{-1ex}
\begin{assumption} \label{as:symmetric}
The graph~$G = (V, E, W) \in \mathcal{G}_n$ has a symmetric weight matrix~$W$, i.e.,~$W = W^T$. 
\end{assumption}
\vspace{-1ex}
We enforce Assumption~\ref{as:symmetric} for simplicity in this work, and we defer analysis 
of the non-symmetric case to a future publication. 
%After representing the networked spreading models through~\eqref{eq:non_dyn_epi_s}-\eqref{eq:non_dyn_epi}, 
We then have the following result to bound the level of penetration
of an epidemic. 
%, i.e., the susceptible proportion within one community at a disease-free equilibrium %of the networked $SIS$ and $SIR$ models 
%as a function of the basic reproduction number $R_0$. %Note that for networked $SIS$ model, the disease-free equilibrium admits zero infection, such that all population are susceptible.
\vspace{-1ex}
\begin{thm}
\label{thm_upper_bound}
    Let~$G \in \mathcal{G}_n$ be given, and
    suppose that a spreading process is modeled either by an $SIS$ or $SIR$ model. 
    Then, for some~$i \in [n]$, there exists a community~$i$ such that the infected proportion $s^*_i$ at disease-free equilibrium is upper bounded via
    $s^*_i\leq\frac{1}{R_0}$.
\end{thm}
\vspace{-1ex}
\noindent\emph{Proof:} See the Appendix~\ref{apdx:proof_equilibrium_bound}. \hfill $\blacksquare$ 

If the nodes in network~$G$ are individuals, then Theorem~\ref{thm_upper_bound} can directly reveal an individual's probability of being infected.
If the nodes are not individuals, then, as discussed in the Introduction, the value of~$R_0$ can reveal sensitive information within~$G$.
Therefore, privacy protections are required that can simultaneously safeguard this information and enable the use of Theorem~\ref{thm_upper_bound} to analyze an epidemic.

%% file: privacy_mechanism_v3.tex
\vspace{-1ex}
\section{Privacy mechanism for $R_0$}\label{sec:privacy_results}
\vspace{-1ex}
%After demonstrating that the basic reproduction number $R_0$ of the overall spreading network can reveal information about entities within the network, 
In this section we develop an input perturbation mechanism to provide differential privacy. 
Specifically, we utilize the bounded Gaussian mechanism~\cite{chen2022bounded} to privatize the next generation matrix $W$.
% In this section, we develop two differentially private mechanisms to privatize the basic reproduction number of the networked spreading dynamics in~\eqref{eq:SIS} and~\eqref{eq:SIR}, as well as their corresponding accuracy bounds. 
%In particular, we consider undirected transmission networks whose transmission matrix $B$ is symmetric. Hence, the constructed matrix of interest, $W=\Gamma^{-1}B$, is also symmetric.
%in undirected graphs whose weight matrices $W$ are symmetric. 
%We leverages two mechanisms.
\vspace{-1.5ex}
\subsection{Input Perturbation Mechanism}
 We start by defining the sensitivity, which quantifies the maximum possible difference between two weighted graphs
  that are adjacent in the sense of Definition~\ref{def:adjacency}.   
\begin{definition}[$L_2$-sensitivity]
Let~$G=(V,E,W) \in \mathcal{G}_n$ and $G'=(V,E,W') \in \mathcal{G}_n$ be adjacent in the sense of Definition~\ref{def:adjacency}. 
Then the $L_2$-sensitivity of the weights, denoted~$\Delta_2w$, is defined as 
%\begin{align*}
    $\Delta_2 w = \max_{G\sim G'} \sqrt{\sum_{i=1}^{n}\sum_{j=1}^n(w_{ij}-w_{ij}')^2}$,
%\end{align*}
where $n=|V|$ is the number of nodes. \hfill $\lozenge$
\end{definition}

%Since the $L_2$ sensitivity is defined by $L_2$ norm of the element-wise entries difference, by the nature of norms we use the fact that $||v||_2\leq||v||_1$ for any vector $v$ to immediately find
%\begin{equation}
%    \Delta_2 w \leq\sum_{i=1}^n\sum_{j=1}^n|w_{ij}-w_{ij}'|\leq k.\label{eq:l2_sensitivity_upper_bound}
%\end{equation}

From Definition~\ref{def:adjacency}, $\Delta_2 w\leq k$. We use this upper bound  to calibrate the variance of noise for privacy protection.

%\begin{mechanism}[Bounded Gaussian mechanism; Solution to Problem~\ref{prob:input_perturbation_design}]\label{mech:bounded_Gaussian_mechanism}
%Fix a probability space $(\Omega,\mathcal{F},\mathbb{P})$. Let $G=(V,E,W)$ be a sensitive graph. Then for $D=(0,\bar{w}]$, the bounded Gaussian mechanism $M_{BG}:D^{n_w}\times \Omega \rightarrow D^{n_w}$ generates independent private weights $\Tilde{w}_{ij}\sim \text{TrunG}(w_{ij},\sigma,0,\Bar{w})$.
%with the probability density function $p_{M_{BG}}$:
%\begin{align}
%    p_{M_{BG}}(\Tilde{w}_{ij}) = \begin{cases}
%        \frac{1}{C_{w_{ij}}(\sigma)} \exp\left(-\frac{(\Tilde{w}_{ij}-w_{ij})^2}{2\sigma^2}\right) & \text{if } \Tilde{w}_{ij}\in D \\
%        0 & \text{otherwise}
%    \end{cases},
%\end{align}
%where $C_{w_{ij}}(\sigma)=\int_D \exp\left(-\frac{(\Tilde{w}_{ij}-w_{ij})^2}{2\sigma^2}\right)d\Tilde{w}_{ij}$ is the normalizing term. 
%The minimal $\sigma=\Omega\left(\sqrt{\frac{2\Bar{w}\Delta_2w+(\Delta_2w)^2}{\epsilon}}\right)$ that satisfies $\epsilon$-differential privacy can be found using \cite[Algorithm 2]{chen2022bounded} by setting $\Delta_2w= k$.\hfill $\lozenge$
%\end{mechanism}

%\begin{mechanism}[Bounded Gaussian mechanism; Solution to Problem~\ref{prob:input_perturbation_design}]\label{mech:bounded_Gaussian_mechanism}
\begin{mechanism}[Bounded Gaussian mechanism]\label{mech:bounded_Gaussian_mechanism}
Fix a probability space $(\Omega,\mathcal{F},\mathbb{P})$. Let $G=(V,E,W) \in \mathcal{G}_n$. Then for $D=(\underline{w}_{ij},\Bar{w}_{ij}]$, the bounded Gaussian mechanism $M_{BG}:D^{n \times n}\times \Omega \rightarrow D^{n \times n}$ generates independent private weights $\Tilde{w}_{ij}\sim \text{TrunG}(w_{ij},\sigma,\underline{w}_{ij},\Bar{w}_{ij})$ for all positive entries $w_{ij}$ on and above the main diagonal of~$W$
(see Section~\ref{ss:setup} for discussion of~$\underline{w}_{ij}$ and $\Bar{w}_{ij}$). 
The entries below the main diagonal mirror the values above it to ensure symmetry.
%The minimal $\sigma^2$ is given by
%$\sigma^2=\Omega\left(\frac{(\sqrt{|E|+|V|}\bar{w}+\frac{1}{2}\Delta_2w)\Delta_2w}{\epsilon}\right)$ 
This mechanism satisfies $\epsilon$-differential privacy if
\begin{equation}
    \sigma^2 \geq \frac{k\left(\frac{k}{2}+\sqrt{\sum_{i=1}^n\sum_{j=i}^n(\bar{w}_{ij}-\underline{w}_{ij})^2\cdot\mathbf{1}_{\mathbb{R}_{> 0}}(w_{ij})}\right)}{\epsilon-\log(\Delta C(\sigma,c))},\label{ineq:sigma}
\end{equation}
where 
%\begin{equation}
    $\Delta C(\sigma,c) = \frac{\Phi\left(\frac{\bar{w}_{ij}-\underline{w}_{ij}-c_{ij}}{\sigma}\right)-\Phi\left(\frac{-c_{ij}}{\sigma}\right)}{\Phi\left(\frac{\bar{w}_{ij}-\underline{w}_{ij}}{\sigma}\right)-\Phi\left(0\right)}$
%\end{equation}
and $c\in\mathbb{R}^{n\times n}\geq 0$ is an upper triangular matrix with $c_{ij}>0$ if and only if $w_{ij}>0$ for all $i,j\in[n]$. Matrix $c$ can be found by solving the optimization problem in~\cite[(3.3)]{chen2022bounded}. 
%that ensures $\epsilon$-differential privacy can be found using \cite[Algorithm 2]{chen2022bounded} with $\Delta_2w=k$.
\hfill $\lozenge$
\end{mechanism}

%When the privacy parameter $\epsilon$ increases, the parameter $\sigma$ in the bounded Gaussian mechanism decreases. And when $\epsilon\rightarrow 0$, we have $\sigma\rightarrow\infty$, and $\text{TrunG}(w_{ij},\sigma,0,\Bar{w})$ converges to $\text{Uniform}(0,\Bar{w})$ for all $w_{ij}$. 
\begin{remark}\label{rmk:implication_epsilon}
    The minimal value of $\sigma$ that satisfies~\eqref{ineq:sigma} can be found using \cite[Algorithm 2]{chen2022bounded}. Meanwhile, \eqref{ineq:sigma} implies that a larger $\epsilon$ gives weaker privacy and leads to a smaller $\sigma$.
    \end{remark}
\vspace{-1ex}
\begin{remark}
The bounded Gaussian mechanism does not add noise to any weight~$w_{ij}=0$. 
Such a weight indicates that there is no edge between nodes~$i$ and~$j$, and thus
the bounded Gaussian mechanism does not alter the presence or absence of an edge in a graph. % by changing a zero weight to a non-zero weight.
\end{remark}
\vspace{-1ex}
Given~$G = (V, E, W)$, and 
suppose the bounded Gaussian mechanism generates an~$\epsilon$-differentially private weights matrix $\Tilde{W}=M_{BG}(W)$. 
Now we can compute a private reproduction number $\Tilde{R}_0$ using the private graph $\Tilde{G}=(V,E,\Tilde{W})$ 
by using $\Tilde{R}_0 = \rho(\Tilde{W})$.
The private reproduction number $\Tilde{R}_0$ provides~$W$ with the same level of privacy protection, $\epsilon$, since differential privacy is immune to post-processing~\cite{dwork2014algorithmic}
and the computation of~$R_0$ simply post-processes the private matrix~$\tilde{W}$.
The accuracy of $\Tilde{R}_0$ is quantified next.
%\baike{[In theorem 1, we didn't define what is a reproduction number of a graph. We need to define this term in the problem formulation section.]}
%The accuracy of $\Tilde{R}_0$ is quantified next.
%\baike{[It is inaccurate to say a reproduction number of a graph. A graph does not has a reproduction number. The definition of the basic reproduction number is based on a matrix. Here, we treat the matrix as a graph. Or we have to define what is the reproduction number of a graph in the problem formulation section.]}
\begin{thm}\label{thm:expectation_bound_input_perturb}
    Consider a graph~$G = (V, E, W)$ and denote its basic reproduction number by~$R_0 = \rho(W)$.
    Suppose Mechanism~\ref{mech:bounded_Gaussian_mechanism} is applied to~$G$, and for all~$i, j \in [n]$ define
    the constants $\alpha_{ij}=\frac{\underline{w}_{ij}-w_{ij}}{\sigma}$ and $\beta_{ij}=\frac{\Bar{w}_{ij}-w_{ij}}{\sigma}$.     
    Also let~$\tilde{G} = (V, E, \tilde{W})$ denote the privatized form of~$G$
    and denote its basic reproduction number by~$\tilde{R}_0 = \rho(\tilde{W})$. 
    Then the error induced in~$R_0$ by privacy obeys the bounds 
    \begin{align}
        &\mathbb{E}\big[|\Tilde{R}_0-R_0|\big] \leq \sigma \sqrt{n_w-\xi_e}\leq \sigma\sqrt{n_w}\label{eq:abs_expectation_bounds_input_perturb}\\
        %&\mathbb{E}\big[(\Tilde{R}_0-R_0)^2\big]\leq \sigma^2\cdot(n_w-\xi_e)\leq \sigma^2n_w\label{eq:mse_bounds_input_perturb},
        &\text{Var}\big[|\Tilde{R}_0-R_0|\big]\leq \sigma^2\cdot(n_w-\xi_e)\leq \sigma^2n_w\label{eq:mse_bounds_input_perturb},
    \end{align}
    where $n_w$ denotes the number of non-zero entries in the weight matrix $W$ and 
    \begin{multline}\label{eq:term_xi}
        \xi_e = 2\sum_{i=1}^n\sum_{j=i+1}^n\frac{\beta_{ij}\varphi(\beta_{ij})-\alpha_{ij}\varphi(\alpha_{ij})}{\Phi(\beta_{ij})-\Phi(\alpha_{ij})}\cdot\mathbf{1}_{\mathbb{R}_{> 0}}(w_{ij})\\
        +\sum_{i=1}^n\frac{\beta_{ii}\varphi(\beta_{ii})-\alpha_{ii}\varphi(\alpha_{ii})}{\Phi(\beta_{ii})-\Phi(\alpha_{ii})}\cdot\mathbf{1}_{\mathbb{R}_{> 0}}(w_{ii}). 
    \end{multline}
    %where $T=\{x\in\mathbb{R}_{> 0}\}$.
\end{thm}

\emph{Proof:} See the Appendix~\ref{apdx:proof_expectation_bound_input_perturb}. \hfill $\blacksquare$

Recall that in Remark~\ref{rmk:implication_epsilon}, a larger $\epsilon$ indicates a smaller $\sigma$, resulting in both $\mathbb{E}[|\Tilde{R}_0-R_0|]$ and $\text{Var}[|\Tilde{R}_0-R_0|]$ being closer to $0$, which is intuitive. In addition to such qualitative analysis, one can use Theorem~\ref{thm:expectation_bound_input_perturb} to predict error on a graph-by-graph basis. For example, consider a complete graph $G=(V,E,W)$ with $|V|=15$ nodes, $|E|=225$ edges (including self loops), and $w_{ij}=0.25$ for all $i,j \in [15]$. If we set $\bar{w}_{ij} = 0.3$ and $\underline{w}_{ij} = 0.2$ for $i,j\in [15]$, and set $\epsilon=5$ and $k=0.01$, 
then we have $\mathbb{E}[|\Tilde{R}_0-R_0|]\leq 0.43$ and $\text{Var}[|\Tilde{R}_0-R_0|]\leq 0.19$, 
where $R_0=3.75$.
%and error ($11.5\%$ error). 
In this example, 
%compute the average infected individuals generated by a single infected case in this example,
the absolute difference $|\tilde{R}_0-R_0|$ is a random variable whose mean and variance are smaller than~$0.43$ and~$0.19$, respectively. Hence, if we use $\tilde{R}_0$ instead of $R_0$ to conduct epidemic analysis, e.g., 
to estimate
the average number of infected individuals generated by a single infected case, 
the deviation that results from using $\tilde{R}_0$ is not likely to be large.
%in terms of the number of generated infected cases
%is in the range of $[0, 0.43]$.}
In general, the bounds in~\eqref{eq:abs_expectation_bounds_input_perturb} and~\eqref{eq:mse_bounds_input_perturb} describe the distribution of the error $|\Tilde{R}_0-R_0|$ in the worst case, which helps analysts to predict the error
that results from providing a given level of privacy protection $\epsilon$.
%gives an upper bound on the ``cost of privacy", i.e., how much accuracy (in this example $0.43$) one needs to sacrifice to satisfy a given differential privacy guarantee (in this example $\epsilon=5$). 

An appealing feature of differential privacy is that its protections are tunable, and here the parameters~$\epsilon$, $k$, $\bar{w}_{ij}$, and $\underline{w}_{ij}$ can be tuned to
balance privacy and accuracy. 
\vspace{-1ex}
\subsection{Use of $\Tilde{R}_0$ for Epidemic Analysis}
Theorem~\ref{thm_upper_bound} shows that~$R_0$ can be used to 
bound the level of penetration in an epidemic spreading network, though,
given the sensitive information that can be revealed
by~$R_0$, it should be privatized before being shared. 
An epidemic analyst may thus
only have access to the private~$R_0$, and 
the question then naturally arises as to how accurate
Theorem~\ref{thm_upper_bound} is when using 
a private value of~$R_0$. We answer this next.
%
%and by utilizing an input perturbation mechanism to compute private $\Tilde{R}_0$, 
%we quantify the accuracy of the level of penetration when the private $\Tilde{R}_0$ is applied to show the effectiveness of our proposed privacy mechanism.
%We start with the input perturbation mechanism in Mechanism~\ref{mech:bounded_Gaussian_mechanism}.
%In this subsection, we quantify the accuracy of the level of penetration when the private $\Tilde{R}_0$ is applied. We start with the input perturbation mechanism in Mechanism~\ref{mech:bounded_Gaussian_mechanism}.

\begin{thm}\label{thm:level_of_penetration_acc_bound_input_perturb}
    Fix a sensitive graph ${G=(V,E,W) \in \mathcal{G}_n}$ and a privacy parameter~$\epsilon$. 
    Consider also a private graph ${\Tilde{G}=(V,E,\Tilde{W})}$ whose weight matrix $\Tilde{W}=M_{BG}(W)$ is generated by Mechanism~\ref{mech:bounded_Gaussian_mechanism}. 
    For the true reproduction number~$R_0 = \rho(W)$, 
    the private reproduction number ${\Tilde{R}_0=\rho(\Tilde{W})}$, and any $t\in(0,R_0-\xi_p)$ we have
    \begin{align}
        \mathbb{P}\left[\left|\frac{1}{\Tilde{R}_0}-\frac{1}{R_0}\right|<\max\{u_1,u_2\} \right] \geq 1-4\exp(-v^2),
    \end{align}
    where 
    \begin{align}
        u_1 &= \frac{1}{R_0}-\frac{1}{R_0+t+\xi_p}, \,\, 
        u_2 = \frac{1}{R_0-t-\xi_p}-\frac{1}{R_0},\\
        %t &= \sqrt{2\sigma^2\cdot(4.4n+v^2)},\\
        v^2 &= \frac{t^2}{2\sigma^2}-4.4n\\
        \xi_p &= \sigma\cdot \sqrt{\sum_{i=1}^n\sum_{j=1}^n\left(\frac{\varphi\left(\alpha_{ij}\right)-\varphi\left(\beta_{ij}\right)}{\Phi\left(\beta_{ij}\right)-\Phi\left(\alpha_{ij}\right)}\right)^2\cdot\mathbf{1}_{\mathbb{R}_{> 0}}(w_{ij})},
    \end{align}
    where the parameter $\sigma$ is from Mechanism~\ref{mech:bounded_Gaussian_mechanism}. 
\end{thm}
\emph{Proof:} See the Appendix~\ref{apdx:proof_level_of_penetration_acc_bound_input_perturb}. \hfill $\blacksquare$

Recall that Theorem~\ref{thm_upper_bound} states that $\frac{1}{R_0}$ bounds the level of penetration. By using Theorem~\ref{thm:level_of_penetration_acc_bound_input_perturb}, we can characterize the distribution of the difference between the true upper bound on the level of penetration, $\frac{1}{R_0}$, and the private upper bound on the level of penetration,  $\frac{1}{\Tilde{R}_0}$. Hence, the result in Theorem~\ref{thm:level_of_penetration_acc_bound_input_perturb} demonstrates the accuracy of Mechanism~\ref{mech:bounded_Gaussian_mechanism} when using the privatizing graph weights. 

For example, consider a complete graph $G=(V,E,W)$ with $|V|=15$ nodes, $|E|=225$ edges (including self loops), and $w_{ij}=0.25$ for each $i,j \in [15]$. Then, if we set $\bar{w}_{ij}=0.3$ and $\underline{w}_{ij}=0.2$ 
for $i,j \in [15]$, and set privacy parameters $\epsilon=5$ and $k=0.01$, we have $\mathbb{P}\left[\left|\frac{1}{\Tilde{R}_0}-\frac{1}{R_0}\right|<0.054\right]\geq 0.92$, 
which indicates that the deviation of using the private upper bound is smaller than~$0.054$ with high probability (0.92),
and thus~$\tilde{R}_0$ can be used without substantially harming accuracy.

%% file: CaseStudy_real_data.tex
\vspace{-2ex}
\section{Simulations}\label{sec:casestudy}
\begin{figure}[tp]
    \centering
    \smallskip
    \smallskip
    \includegraphics[scale=0.2]{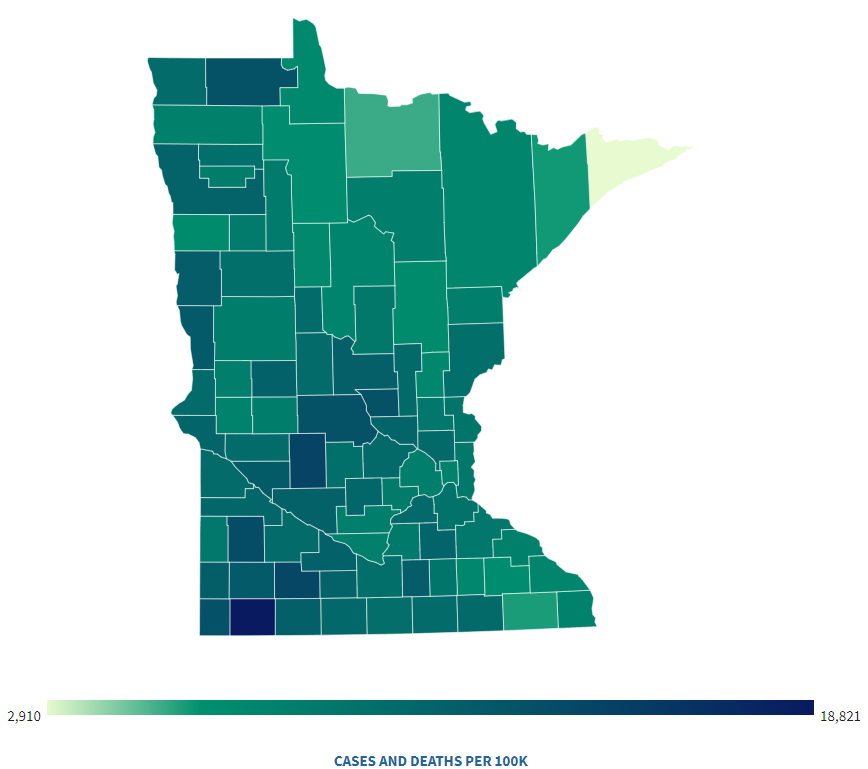}
    \caption{Infection map of Minnesota~\cite{Minnesota_2023}.}
    \label{fig:MN}

\end{figure}

\vspace{-1ex}
In this section, we present simulation results for generating $\Tilde{R}_0$ using Mechanism~\ref{mech:bounded_Gaussian_mechanism}. 
We use a graph $G=(V,E,W)$ to model networked data that estimates the number of trips between Minnesota counties \cite{Minnesota_2023} (shown in~Figure~\ref{fig:MN}) via 
%anonymous 
geolocalization using smartphones~\cite{brooks2023TNSEflows}.
The data provides an estimate of the total number of trips made by individuals between counties 
in Minnesota from March 2020 to December 2020. %\mh{What specifically is the period of time?} 
We choose a weekly time scale in an effort to average out periodic behaviors and use this average to estimate the daily flow of individuals between counties.  
The work in \cite{brooks2023TNSEflows} constructs the asymmetric transmission matrix $B'$ by leveraging the daily flow between two counties, i.e., by setting $b'_{ij}$ as the daily traffic flow from county $i$ to $j$, for all $i$, $j\in [87]$. In order to satisfy Assumption~\ref{as:symmetric}, we set the matrix $B$ with $b_{ij}=b_{ji}=\frac{b'_{ij}+b'_{ji}}{87}$ and $b_{ii} = \frac{|\sum_{i}b'_{ij}-\sum_{j}b'_{ij}|}{87}$ for all $i,j\in[87]$, which results in
$b_{ij}\in [1.172\times10^{-6}, 0.621]$ for all $i,j\in [87]$.
The recovery rate for all~$i \in [87]$ %\mh{I added ``for all~$i \in [87]$.'' Is that correct}
is $\gamma_i=\frac{1}{3}$. Thus, the next generation matrix of~$G$, namely $W=\Gamma^{-1}B$, is symmetric with $|V|=87$ representing Minnesota’s~$87$ counties, and $|E|=3565$ is the number of edges that 
represent travel connections between pairs of counties. The network's
basic reproduction number is $R_0=\rho(W)=3.54$.

% The recovery rate for all~$i \in [87]$ %\mh{I added ``for all~$i \in [87]$.'' Is that correct}
% is $\gamma_i=\frac{1}{3}$, and the network's
% basic reproduction number is $R_0=3.54$.
Through this formulation of $B$ and $W$, the weights in $W$ 
are proportional to the volume of travel between counties, and larger values of an entry $w_{ij}$ indicate a higher volume of travel between counties $i$ and $j$. We classify the weights into three categories, which are low, medium, and high travel flows, which correspond to the weight ranges
$(0,0.01]$, $(0.01,0.1]$, and $(0.1,3]$, respectively. We set the adjacency parameter to $k=0.001$. This choice of $k$ is because over half of the entries in the weight matrix $W$ are much smaller than $k$, indicating that this choice of~$k$ certainly fulfills our objective of protecting individuals. 
In fact, in more than half of the entries of~$W$, it simultaneously protects \emph{all} individuals whose travel is encoded in that entry. 
In our simulations, we generated~$100$ private graphs for each $\epsilon\in[5,20]$ using Mechanism~\ref{mech:bounded_Gaussian_mechanism} on~$G$.
%\mh{Do we do~$100$ for each~$\epsilon$?}

\begin{figure}[tp]
    \centering
    % This file was created with tikzplotlib v0.10.1.
    \begin{tikzpicture}
    
    \definecolor{darkgray176}{RGB}{176,176,176}
    \definecolor{darkorange25512714}{RGB}{255,127,14}
    
    \begin{axis}[
    width=\columnwidth, 
    height=3.5cm,
    tick align=outside,
    tick pos=left,
    x grid style={darkgray176},
    xlabel={\footnotesize Privacy Parameter \(\displaystyle \epsilon\)},
    xmajorgrids,
    xmin=0.5, xmax=16.5,
    xtick style={color=black},
    xtick={2,4,6,8,10,12,14,16},
    xticklabels={6,8,10,12,14,16,18,20},
    y grid style={darkgray176},
    ylabel={\footnotesize\(\displaystyle \left|\frac{1}{\tilde{R}_0}-\frac{1}{R_0}\right|\)},
    ymajorgrids,
    ymin=0.0124391267003456, ymax=0.0404893038203972,
    ytick style={color=black}
    ]
    \addplot [very thick, red, mark=*, mark size=1, mark options={solid}]
    table {%
    1 0.0314604841444847
    2 0.0291206293692653
    3 0.0280259782897483
    4 0.0264255889702425
    5 0.0257165096254672
    6 0.0244916276205982
    7 0.0239008878462599
    8 0.0230444877316844
    9 0.0224887876992386
    10 0.0222167195338946
    11 0.0215740758528439
    12 0.0213227164390474
    13 0.0211237248290561
    14 0.0205676525522552
    15 0.0205883341432593
    16 0.0197334021569073
    };
    \addplot [black]
    table {%
    0.75 0.0285839180497966
    1.25 0.0285839180497966
    1.25 0.0350063775945853
    0.75 0.0350063775945853
    0.75 0.0285839180497966
    };
    \addplot [black]
    table {%
    1 0.0285839180497966
    1 0.0190761288843026
    };
    \addplot [black]
    table {%
    1 0.0350063775945853
    1 0.0392142957694857
    };
    \addplot [black]
    table {%
    0.875 0.0190761288843026
    1.125 0.0190761288843026
    };
    \addplot [black]
    table {%
    0.875 0.0392142957694857
    1.125 0.0392142957694857
    };
    \addplot [black]
    table {%
    1.75 0.0268329610033421
    2.25 0.0268329610033421
    2.25 0.0310784170855881
    1.75 0.0310784170855881
    1.75 0.0268329610033421
    };
    \addplot [black]
    table {%
    2 0.0268329610033421
    2 0.0207185777260043
    };
    \addplot [black]
    table {%
    2 0.0310784170855881
    2 0.0373754403938341
    };
    \addplot [black]
    table {%
    1.875 0.0207185777260043
    2.125 0.0207185777260043
    };
    \addplot [black]
    table {%
    1.875 0.0373754403938341
    2.125 0.0373754403938341
    };
    \addplot [black]
    table {%
    2.75 0.0251723481993568
    3.25 0.0251723481993568
    3.25 0.0310307221323786
    2.75 0.0310307221323786
    2.75 0.0251723481993568
    };
    \addplot [black]
    table {%
    3 0.0251723481993568
    3 0.0210112885641102
    };
    \addplot [black]
    table {%
    3 0.0310307221323786
    3 0.0391416039403607
    };
    \addplot [black]
    table {%
    2.875 0.0210112885641102
    3.125 0.0210112885641102
    };
    \addplot [black]
    table {%
    2.875 0.0391416039403607
    3.125 0.0391416039403607
    };
    \addplot [black]
    table {%
    3.75 0.0239048141220161
    4.25 0.0239048141220161
    4.25 0.0289656162695729
    3.75 0.0289656162695729
    3.75 0.0239048141220161
    };
    \addplot [black]
    table {%
    4 0.0239048141220161
    4 0.0169680738000734
    };
    \addplot [black]
    table {%
    4 0.0289656162695729
    4 0.0333972887003516
    };
    \addplot [black]
    table {%
    3.875 0.0169680738000734
    4.125 0.0169680738000734
    };
    \addplot [black]
    table {%
    3.875 0.0333972887003516
    4.125 0.0333972887003516
    };
    \addplot [black]
    table {%
    4.75 0.0235055645418397
    5.25 0.0235055645418397
    5.25 0.0282215120243021
    4.75 0.0282215120243021
    4.75 0.0235055645418397
    };
    \addplot [black]
    table {%
    5 0.0235055645418397
    5 0.0174697474108926
    };
    \addplot [black]
    table {%
    5 0.0282215120243021
    5 0.0342685766115039
    };
    \addplot [black]
    table {%
    4.875 0.0174697474108926
    5.125 0.0174697474108926
    };
    \addplot [black]
    table {%
    4.875 0.0342685766115039
    5.125 0.0342685766115039
    };
    \addplot [black]
    table {%
    5.75 0.0222876193639091
    6.25 0.0222876193639091
    6.25 0.0268405671462653
    5.75 0.0268405671462653
    5.75 0.0222876193639091
    };
    \addplot [black]
    table {%
    6 0.0222876193639091
    6 0.0187422416122441
    };
    \addplot [black]
    table {%
    6 0.0268405671462653
    6 0.0303283868297036
    };
    \addplot [black]
    table {%
    5.875 0.0187422416122441
    6.125 0.0187422416122441
    };
    \addplot [black]
    table {%
    5.875 0.0303283868297036
    6.125 0.0303283868297036
    };
    \addplot [black]
    table {%
    6.75 0.0216699839416594
    7.25 0.0216699839416594
    7.25 0.0258629965217795
    6.75 0.0258629965217795
    6.75 0.0216699839416594
    };
    \addplot [black]
    table {%
    7 0.0216699839416594
    7 0.0168344654735046
    };
    \addplot [black]
    table {%
    7 0.0258629965217795
    7 0.0296082225956676
    };
    \addplot [black]
    table {%
    6.875 0.0168344654735046
    7.125 0.0168344654735046
    };
    \addplot [black]
    table {%
    6.875 0.0296082225956676
    7.125 0.0296082225956676
    };
    \addplot [black]
    table {%
    7.75 0.0213154898959119
    8.25 0.0213154898959119
    8.25 0.0248040922747366
    7.75 0.0248040922747366
    7.75 0.0213154898959119
    };
    \addplot [black]
    table {%
    8 0.0213154898959119
    8 0.0161033568567325
    };
    \addplot [black]
    table {%
    8 0.0248040922747366
    8 0.0291238329364835
    };
    \addplot [black]
    table {%
    7.875 0.0161033568567325
    8.125 0.0161033568567325
    };
    \addplot [black]
    table {%
    7.875 0.0291238329364835
    8.125 0.0291238329364835
    };
    \addplot [black]
    table {%
    8.75 0.0207610856622761
    9.25 0.0207610856622761
    9.25 0.0243718226194834
    8.75 0.0243718226194834
    8.75 0.0207610856622761
    };
    \addplot [black]
    table {%
    9 0.0207610856622761
    9 0.0163199083443915
    };
    \addplot [black]
    table {%
    9 0.0243718226194834
    9 0.0296472092344406
    };
    \addplot [black]
    table {%
    8.875 0.0163199083443915
    9.125 0.0163199083443915
    };
    \addplot [black]
    table {%
    8.875 0.0296472092344406
    9.125 0.0296472092344406
    };
    \addplot [black]
    table {%
    9.75 0.0205267179672499
    10.25 0.0205267179672499
    10.25 0.0238752166456573
    9.75 0.0238752166456573
    9.75 0.0205267179672499
    };
    \addplot [black]
    table {%
    10 0.0205267179672499
    10 0.0162812104868474
    };
    \addplot [black]
    table {%
    10 0.0238752166456573
    10 0.0288570978068192
    };
    \addplot [black]
    table {%
    9.875 0.0162812104868474
    10.125 0.0162812104868474
    };
    \addplot [black]
    table {%
    9.875 0.0288570978068192
    10.125 0.0288570978068192
    };
    \addplot [black]
    table {%
    10.75 0.0197382400728186
    11.25 0.0197382400728186
    11.25 0.0235813342860321
    10.75 0.0235813342860321
    10.75 0.0197382400728186
    };
    \addplot [black]
    table {%
    11 0.0197382400728186
    11 0.0160788380421643
    };
    \addplot [black]
    table {%
    11 0.0235813342860321
    11 0.0289297812011922
    };
    \addplot [black]
    table {%
    10.875 0.0160788380421643
    11.125 0.0160788380421643
    };
    \addplot [black]
    table {%
    10.875 0.0289297812011922
    11.125 0.0289297812011922
    };
    \addplot [black]
    table {%
    11.75 0.019399678207772
    12.25 0.019399678207772
    12.25 0.023404458764482
    11.75 0.023404458764482
    11.75 0.019399678207772
    };
    \addplot [black]
    table {%
    12 0.019399678207772
    12 0.0140108754806805
    };
    \addplot [black]
    table {%
    12 0.023404458764482
    12 0.0278545306600819
    };
    \addplot [black]
    table {%
    11.875 0.0140108754806805
    12.125 0.0140108754806805
    };
    \addplot [black]
    table {%
    11.875 0.0278545306600819
    12.125 0.0278545306600819
    };
    \addplot [black]
    table {%
    12.75 0.019565451729339
    13.25 0.019565451729339
    13.25 0.0226405979426059
    12.75 0.0226405979426059
    12.75 0.019565451729339
    };
    \addplot [black]
    table {%
    13 0.019565451729339
    13 0.0168854366887136
    };
    \addplot [black]
    table {%
    13 0.0226405979426059
    13 0.0272282786186605
    };
    \addplot [black]
    table {%
    12.875 0.0168854366887136
    13.125 0.0168854366887136
    };
    \addplot [black]
    table {%
    12.875 0.0272282786186605
    13.125 0.0272282786186605
    };
    \addplot [black]
    table {%
    13.75 0.019011058676533
    14.25 0.019011058676533
    14.25 0.0217414920261718
    13.75 0.0217414920261718
    13.75 0.019011058676533
    };
    \addplot [black]
    table {%
    14 0.019011058676533
    14 0.0150530057916732
    };
    \addplot [black]
    table {%
    14 0.0217414920261718
    14 0.0257399563919842
    };
    \addplot [black]
    table {%
    13.875 0.0150530057916732
    14.125 0.0150530057916732
    };
    \addplot [black]
    table {%
    13.875 0.0257399563919842
    14.125 0.0257399563919842
    };
    \addplot [black]
    table {%
    14.75 0.0189173077465174
    15.25 0.0189173077465174
    15.25 0.022580802591436
    14.75 0.022580802591436
    14.75 0.0189173077465174
    };
    \addplot [black]
    table {%
    15 0.0189173077465174
    15 0.014147318135587
    };
    \addplot [black]
    table {%
    15 0.022580802591436
    15 0.0267374017667011
    };
    \addplot [black]
    table {%
    14.875 0.014147318135587
    15.125 0.014147318135587
    };
    \addplot [black]
    table {%
    14.875 0.0267374017667011
    15.125 0.0267374017667011
    };
    \addplot [black]
    table {%
    15.75 0.0181319885560069
    16.25 0.0181319885560069
    16.25 0.0210785145623861
    15.75 0.0210785145623861
    15.75 0.0181319885560069
    };
    \addplot [black]
    table {%
    16 0.0181319885560069
    16 0.013714134751257
    };
    \addplot [black]
    table {%
    16 0.0210785145623861
    16 0.024440257929303
    };
    \addplot [black]
    table {%
    15.875 0.013714134751257
    16.125 0.013714134751257
    };
    \addplot [black]
    table {%
    15.875 0.024440257929303
    16.125 0.024440257929303
    };
    \addplot [darkorange25512714]
    table {%
    0.75 0.0320864520747616
    1.25 0.0320864520747616
    };
    \addplot [darkorange25512714]
    table {%
    1.75 0.0287540332950623
    2.25 0.0287540332950623
    };
    \addplot [darkorange25512714]
    table {%
    2.75 0.028132230195083
    3.25 0.028132230195083
    };
    \addplot [darkorange25512714]
    table {%
    3.75 0.026382450121568
    4.25 0.026382450121568
    };
    \addplot [darkorange25512714]
    table {%
    4.75 0.0251999785969112
    5.25 0.0251999785969112
    };
    \addplot [darkorange25512714]
    table {%
    5.75 0.0242738471467939
    6.25 0.0242738471467939
    };
    \addplot [darkorange25512714]
    table {%
    6.75 0.0239018273063291
    7.25 0.0239018273063291
    };
    \addplot [darkorange25512714]
    table {%
    7.75 0.0232501399967374
    8.25 0.0232501399967374
    };
    \addplot [darkorange25512714]
    table {%
    8.75 0.0228109927308141
    9.25 0.0228109927308141
    };
    \addplot [darkorange25512714]
    table {%
    9.75 0.0217655165419853
    10.25 0.0217655165419853
    };
    \addplot [darkorange25512714]
    table {%
    10.75 0.021270643957789
    11.25 0.021270643957789
    };
    \addplot [darkorange25512714]
    table {%
    11.75 0.0215178412577304
    12.25 0.0215178412577304
    };
    \addplot [darkorange25512714]
    table {%
    12.75 0.0207583064670282
    13.25 0.0207583064670282
    };
    \addplot [darkorange25512714]
    table {%
    13.75 0.0204744335406929
    14.25 0.0204744335406929
    };
    \addplot [darkorange25512714]
    table {%
    14.75 0.0206238674486739
    15.25 0.0206238674486739
    };
    \addplot [darkorange25512714]
    table {%
    15.75 0.0196631823412924
    16.25 0.0196631823412924
    };
    \end{axis}
    
    \end{tikzpicture}
    
    \caption{The value of $\left|\frac{1}{\Tilde{R}_0}-\frac{1}{R_0}\right|$ as a function of the privacy parameter $\epsilon$ given $\frac{1}{R_0}=0.283$. Smaller values of  $\epsilon$
    correspond to stronger privacy.}
    \label{fig:abs_error_pene}
\end{figure}
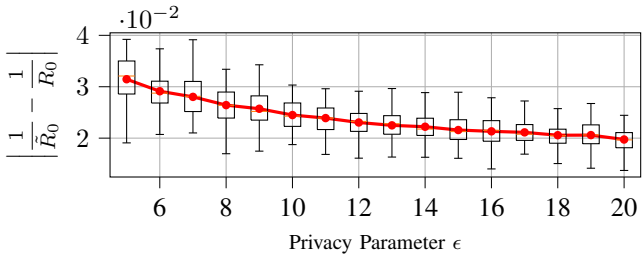

We compute and plot 
%the empirical average of 
%the absolute differences $|\Tilde{R}_0-R_0|$ for each $\epsilon\in[5,20]$, 
the absolute differences $|\Tilde{R}_0-R_0|$ and $\left|\frac{1}{\Tilde{R}_0}-\frac{1}{R_0}\right|$ for each $\epsilon\in[5,20]$, which are shown in Figures~\ref{fig:abs_avg_acc_loss_vs_epsilon} and~\ref{fig:abs_error_pene}, respectively. 
%Figure~\ref{fig:abs_avg_acc_loss_vs_epsilon} plots these values to illustrate the effect of privacy on the accuracy of the value of $\tilde{R}_0$. 
Recall from Remark~\ref{rmk:implication_epsilon} that higher values of the privacy parameter $\epsilon$ imply weaker privacy, and the simulation 
results confirm that weaker privacy guarantees result in smaller errors. 
For all values of $\epsilon\in[5,20]$, the empirical average of $|\tilde{R}_0-R_0|$ is small (between $0.27$ and $0.45$, incurring errors from $7.6\%$ to $12.7\%$ on average) compared to the true value $R_0=3.54$. 
Similarly, the empirical average of $\left|\frac{1}{\Tilde{R}_0}-\frac{1}{R_0}\right|$ is from $0.019$ to $0.031$, incurring errors from $7.0\%$ to $11.2\%$.
Additionally, both the values of $|\Tilde{R}_0-R_0|$ and $\left|\frac{1}{\Tilde{R}_0}-\frac{1}{R_0}\right|$ are concentrated around their empirical averages. These simulation results demonstrate that $\tilde{R}_0$ maintains enough accuracy under privacy to
enable useful analyses alongside protecting information.

%a strong performance of the proposed input perturbation mechanism. 
%\begin{figure}
%    \centering
%    \begin{tikzpicture}
%
%    \begin{axis}[
%    width=\columnwidth, 
%    height=4cm,
%    legend cell align={left},
%    legend style={fill opacity=0.8, draw opacity=1, text opacity=1, draw=white!80!black},
%    tick align=outside,
%    tick pos=left,
%    x grid style={white!69.0196078431373!black},
%    xlabel={\footnotesize Privacy Parameter \(\displaystyle \epsilon\)},
%    xmajorgrids,
%    xmin=4.25, xmax=20.75,
%    xtick style={color=black},
%    y grid style={white!69.0196078431373!black},
%    ylabel={\footnotesize  Avg. of \(\displaystyle |\tilde{R}_0-R_0|\)},
%    ymajorgrids,
%    ymin=0.256326124273972, ymax=0.452117043519429,
%    ytick style={color=black}
%    ]
%    \addplot [very thick, red, mark=*, mark size=2, mark options={solid}]
%    table {%
%    5 0.443217456280999
%    6 0.406047123499859
%    7 0.389256849402652
%    8 0.364589356251373
%    9 0.353820175370833
%    10 0.335167834018892
%    11 0.326387127529473
%    12 0.313748237352131
%    13 0.305493307646673
%    14 0.301489187217332
%    15 0.291989236981117
%    16 0.288397235293848
%    17 0.285376996072965
%    18 0.277232014488117
%    19 0.277636422638081
%    20 0.265225711512402
%    };
    %\addlegendentry{Input Pertubation}
%    \end{axis}
%   
%    \end{tikzpicture}
%    
%    \caption{Empirical average of $|\tilde{R}_0-R_0|$ as a function of the privacy parameter $\epsilon$ given $R_0=3.54$. Smaller values of  $\epsilon$
%    correspond to stronger privacy.}
    %\label{fig:abs_avg_acc_loss_vs_epsilon}
%\end{figure}

\begin{figure}[tp]
    \centering
    \smallskip
    \smallskip
    % This file was created with tikzplotlib v0.10.1.
    \begin{tikzpicture}
    
    \definecolor{darkgray176}{RGB}{176,176,176}
    \definecolor{darkorange25512714}{RGB}{255,127,14}
    \definecolor{lightgray204}{RGB}{204,204,204}
    
    \begin{axis}[
    width=\columnwidth, 
    height=3.5cm,
    legend style={fill opacity=0.8, draw opacity=1, text opacity=1, draw=lightgray204},
    tick align=outside,
    tick pos=left,
    x grid style={darkgray176},
    xlabel={\footnotesize Privacy Parameter \(\displaystyle \epsilon\)},
    xmajorgrids,
    xmin=0.5, xmax=16.5,
    xtick style={color=black},
    xtick={2,4,6,8,10,12,14,16},
    xticklabels={6,8,10,12,14,16,18,20},
    y grid style={darkgray176},
    ylabel={\footnotesize  \(\displaystyle |\tilde{R}_0-R_0|\)},
    ymajorgrids,
    ymin=0.160523380585099, ymax=0.58777096565606,
    ytick style={color=black}
    ]
    \addplot [very thick, red, mark=*, mark size=1, mark options={solid}, forget plot]
    table {%
    1 0.443217456280999
    2 0.406047123499859
    3 0.389256849402652
    4 0.364589356251373
    5 0.353820175370833
    6 0.335167834018892
    7 0.326387127529473
    8 0.313748237352131
    9 0.305493307646673
    10 0.301489187217332
    11 0.291989236981117
    12 0.288397235293848
    13 0.285376996072965
    14 0.277232014488117
    15 0.277636422638081
    16 0.265225711512402
    };
    \addplot [black, forget plot]
    table {%
    0.75 0.396970381907883
    1.25 0.396970381907883
    1.25 0.498754949543313
    0.75 0.498754949543313
    0.75 0.396970381907883
    };
    \addplot [black, forget plot]
    table {%
    1 0.396970381907883
    1 0.255383655021477
    };
    \addplot [black, forget plot]
    table {%
    1 0.498754949543313
    1 0.568350620880107
    };
    \addplot [black, forget plot]
    table {%
    0.875 0.255383655021477
    1.125 0.255383655021477
    };
    \addplot [black, forget plot]
    table {%
    0.875 0.568350620880107
    1.125 0.568350620880107
    };
    \addplot [black, forget plot]
    table {%
    1.75 0.370106250940076
    2.25 0.370106250940076
    2.25 0.435887356196531
    1.75 0.435887356196531
    1.75 0.370106250940076
    };
    \addplot [black, forget plot]
    table {%
    2 0.370106250940076
    2 0.279108991781367
    };
    \addplot [black, forget plot]
    table {%
    2 0.435887356196531
    2 0.53223682005647
    };
    \addplot [black, forget plot]
    table {%
    1.875 0.279108991781367
    2.125 0.279108991781367
    };
    \addplot [black, forget plot]
    table {%
    1.875 0.53223682005647
    2.125 0.53223682005647
    };
    \addplot [black, forget plot]
    table {%
    2.75 0.344965343549065
    3.25 0.344965343549065
    3.25 0.435136106586347
    2.75 0.435136106586347
    2.75 0.344965343549065
    };
    \addplot [black, forget plot]
    table {%
    3 0.344965343549065
    3 0.283368459280928
    };
    \addplot [black, forget plot]
    table {%
    3 0.435136106586347
    3 0.567127966363665
    };
    \addplot [black, forget plot]
    table {%
    2.875 0.283368459280928
    3.125 0.283368459280928
    };
    \addplot [black, forget plot]
    table {%
    2.875 0.567127966363665
    3.125 0.567127966363665
    };
    \addplot [black, forget plot]
    table {%
    3.75 0.32599223052212
    4.25 0.32599223052212
    4.25 0.402875779138951
    3.75 0.402875779138951
    3.75 0.32599223052212
    };
    \addplot [black, forget plot]
    table {%
    4 0.32599223052212
    4 0.225361871048855
    };
    \addplot [black, forget plot]
    table {%
    4 0.402875779138951
    4 0.472761915792567
    };
    \addplot [black, forget plot]
    table {%
    3.875 0.225361871048855
    4.125 0.225361871048855
    };
    \addplot [black, forget plot]
    table {%
    3.875 0.472761915792567
    4.125 0.472761915792567
    };
    \addplot [black, forget plot]
    table {%
    4.75 0.320054456038038
    5.25 0.320054456038038
    5.25 0.39137983892744
    4.75 0.39137983892744
    4.75 0.320054456038038
    };
    \addplot [black, forget plot]
    table {%
    5 0.320054456038038
    5 0.232463215222873
    };
    \addplot [black, forget plot]
    table {%
    5 0.39137983892744
    5 0.486794789335856
    };
    \addplot [black, forget plot]
    table {%
    4.875 0.232463215222873
    5.125 0.232463215222873
    };
    \addplot [black, forget plot]
    table {%
    4.875 0.486794789335856
    5.125 0.486794789335856
    };
    \addplot [black, forget plot]
    table {%
    5.75 0.302053138763303
    6.25 0.302053138763303
    6.25 0.370222116412248
    5.75 0.370222116412248
    5.75 0.302053138763303
    };
    \addplot [black, forget plot]
    table {%
    6 0.302053138763303
    6 0.25059668942419
    };
    \addplot [black, forget plot]
    table {%
    6 0.370222116412248
    6 0.42410524372203
    };
    \addplot [black, forget plot]
    table {%
    5.875 0.25059668942419
    6.125 0.25059668942419
    };
    \addplot [black, forget plot]
    table {%
    5.875 0.42410524372203
    6.125 0.42410524372203
    };
    \addplot [black, forget plot]
    table {%
    6.75 0.292989248774694
    7.25 0.292989248774694
    7.25 0.355381984744888
    6.75 0.355381984744888
    6.75 0.292989248774694
    };
    \addplot [black, forget plot]
    table {%
    7 0.292989248774694
    7 0.223475118006613
    };
    \addplot [black, forget plot]
    table {%
    7 0.355381984744888
    7 0.412857951177741
    };
    \addplot [black, forget plot]
    table {%
    6.875 0.223475118006613
    7.125 0.223475118006613
    };
    \addplot [black, forget plot]
    table {%
    6.875 0.412857951177741
    7.125 0.412857951177741
    };
    \addplot [black, forget plot]
    table {%
    7.75 0.287805235174492
    8.25 0.287805235174492
    8.25 0.339433848361063
    7.75 0.339433848361063
    7.75 0.287805235174492
    };
    \addplot [black, forget plot]
    table {%
    8 0.287805235174492
    8 0.213184199534575
    };
    \addplot [black, forget plot]
    table {%
    8 0.339433848361063
    8 0.405328801896438
    };
    \addplot [black, forget plot]
    table {%
    7.875 0.213184199534575
    8.125 0.213184199534575
    };
    \addplot [black, forget plot]
    table {%
    7.875 0.405328801896438
    8.125 0.405328801896438
    };
    \addplot [black, forget plot]
    table {%
    8.75 0.279727279049816
    9.25 0.279727279049816
    9.25 0.332961197518951
    8.75 0.332961197518951
    8.75 0.279727279049816
    };
    \addplot [black, forget plot]
    table {%
    9 0.279727279049816
    9 0.216226447863134
    };
    \addplot [black, forget plot]
    table {%
    9 0.332961197518951
    9 0.396653309326691
    };
    \addplot [black, forget plot]
    table {%
    8.875 0.216226447863134
    9.125 0.216226447863134
    };
    \addplot [black, forget plot]
    table {%
    8.875 0.396653309326691
    9.125 0.396653309326691
    };
    \addplot [black, forget plot]
    table {%
    9.75 0.276322852153267
    10.25 0.276322852153267
    10.25 0.325551420898098
    9.75 0.325551420898098
    9.75 0.276322852153267
    };
    \addplot [black, forget plot]
    table {%
    10 0.276322852153267
    10 0.215682434083571
    };
    \addplot [black, forget plot]
    table {%
    10 0.325551420898098
    10 0.385216309097797
    };
    \addplot [black, forget plot]
    table {%
    9.875 0.215682434083571
    10.125 0.215682434083571
    };
    \addplot [black, forget plot]
    table {%
    9.875 0.385216309097797
    10.125 0.385216309097797
    };
    \addplot [black, forget plot]
    table {%
    10.75 0.26491232622746
    11.25 0.26491232622746
    11.25 0.321179932153119
    10.75 0.321179932153119
    10.75 0.26491232622746
    };
    \addplot [black, forget plot]
    table {%
    11 0.26491232622746
    11 0.212840055249059
    };
    \addplot [black, forget plot]
    table {%
    11 0.321179932153119
    11 0.402320597248395
    };
    \addplot [black, forget plot]
    table {%
    10.875 0.212840055249059
    11.125 0.212840055249059
    };
    \addplot [black, forget plot]
    table {%
    10.875 0.402320597248395
    11.125 0.402320597248395
    };
    \addplot [black, forget plot]
    table {%
    11.75 0.260033974838706
    12.25 0.260033974838706
    12.25 0.318553863132682
    11.75 0.318553863132682
    11.75 0.260033974838706
    };
    \addplot [black, forget plot]
    table {%
    12 0.260033974838706
    12 0.184040068825741
    };
    \addplot [black, forget plot]
    table {%
    12 0.318553863132682
    12 0.385734902249098
    };
    \addplot [black, forget plot]
    table {%
    11.875 0.184040068825741
    12.125 0.184040068825741
    };
    \addplot [black, forget plot]
    table {%
    11.875 0.385734902249098
    12.125 0.385734902249098
    };
    \addplot [black, forget plot]
    table {%
    12.75 0.262421131059546
    13.25 0.262421131059546
    13.25 0.307252856493859
    12.75 0.307252856493859
    12.75 0.262421131059546
    };
    \addplot [black, forget plot]
    table {%
    13 0.262421131059546
    13 0.196970895985927
    };
    \addplot [black, forget plot]
    table {%
    13 0.307252856493859
    13 0.370890248047288
    };
    \addplot [black, forget plot]
    table {%
    12.875 0.196970895985927
    13.125 0.196970895985927
    };
    \addplot [black, forget plot]
    table {%
    12.875 0.370890248047288
    13.125 0.370890248047288
    };
    \addplot [black, forget plot]
    table {%
    13.75 0.254450020566002
    14.25 0.254450020566002
    14.25 0.294035899458172
    13.75 0.294035899458172
    13.75 0.254450020566002
    };
    \addplot [black, forget plot]
    table {%
    14 0.254450020566002
    14 0.19849798887389
    };
    \addplot [black, forget plot]
    table {%
    14 0.294035899458172
    14 0.337187777742451
    };
    \addplot [black, forget plot]
    table {%
    13.875 0.19849798887389
    14.125 0.19849798887389
    };
    \addplot [black, forget plot]
    table {%
    13.875 0.337187777742451
    14.125 0.337187777742451
    };
    \addplot [black, forget plot]
    table {%
    14.75 0.253105169746101
    15.25 0.253105169746101
    15.25 0.306371019789308
    14.75 0.306371019789308
    14.75 0.253105169746101
    };
    \addplot [black, forget plot]
    table {%
    15 0.253105169746101
    15 0.18592661971145
    };
    \addplot [black, forget plot]
    table {%
    15 0.306371019789308
    15 0.368650650074366
    };
    \addplot [black, forget plot]
    table {%
    14.875 0.18592661971145
    15.125 0.18592661971145
    };
    \addplot [black, forget plot]
    table {%
    14.875 0.368650650074366
    15.125 0.368650650074366
    };
    \addplot [black, forget plot]
    table {%
    15.75 0.24187863978845
    16.25 0.24187863978845
    16.25 0.284348340635396
    15.75 0.284348340635396
    15.75 0.24187863978845
    };
    \addplot [black, forget plot]
    table {%
    16 0.24187863978845
    16 0.179943725361051
    };
    \addplot [black, forget plot]
    table {%
    16 0.284348340635396
    16 0.333984304663028
    };
    \addplot [black, forget plot]
    table {%
    15.875 0.179943725361051
    16.125 0.179943725361051
    };
    \addplot [black, forget plot]
    table {%
    15.875 0.333984304663028
    16.125 0.333984304663028
    };
    \addplot [darkorange25512714, forget plot]
    table {%
    0.75 0.451835576458371
    1.25 0.451835576458371
    };
    \addplot [darkorange25512714, forget plot]
    table {%
    1.75 0.399600223878209
    2.25 0.399600223878209
    };
    \addplot [darkorange25512714, forget plot]
    table {%
    2.75 0.390005129934939
    3.25 0.390005129934939
    };
    \addplot [darkorange25512714, forget plot]
    table {%
    3.75 0.363255151374201
    4.25 0.363255151374201
    };
    \addplot [darkorange25512714, forget plot]
    table {%
    4.75 0.345381156047006
    5.25 0.345381156047006
    };
    \addplot [darkorange25512714, forget plot]
    table {%
    5.75 0.331497052221558
    6.25 0.331497052221558
    };
    \addplot [darkorange25512714, forget plot]
    table {%
    6.75 0.325947748942491
    7.25 0.325947748942491
    };
    \addplot [darkorange25512714, forget plot]
    table {%
    7.75 0.316265279937069
    8.25 0.316265279937069
    };
    \addplot [darkorange25512714, forget plot]
    table {%
    8.75 0.309767983541011
    9.25 0.309767983541011
    };
    \addplot [darkorange25512714, forget plot]
    table {%
    9.75 0.29438781264089
    10.25 0.29438781264089
    };
    \addplot [darkorange25512714, forget plot]
    table {%
    10.75 0.287150539534476
    11.25 0.287150539534476
    };
    \addplot [darkorange25512714, forget plot]
    table {%
    11.75 0.290762241582437
    12.25 0.290762241582437
    };
    \addplot [darkorange25512714, forget plot]
    table {%
    12.75 0.27968656038854
    13.25 0.27968656038854
    };
    \addplot [darkorange25512714, forget plot]
    table {%
    13.75 0.275563531125133
    14.25 0.275563531125133
    };
    \addplot [darkorange25512714, forget plot]
    table {%
    14.75 0.277732823225492
    15.25 0.277732823225492
    };
    \addplot [darkorange25512714, forget plot]
    table {%
    15.75 0.26382979395815
    16.25 0.26382979395815
    };
    \end{axis}
    
    \end{tikzpicture}
    \caption{The value of $|\tilde{R}_0-R_0|$ as a function of the privacy parameter $\epsilon$ given $R_0=3.54$. Smaller values of  $\epsilon$
    correspond to stronger privacy.}

\label{fig:abs_avg_acc_loss_vs_epsilon}
\end{figure}
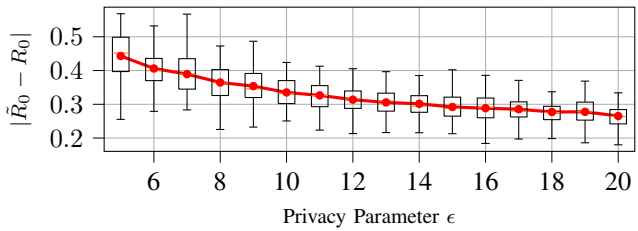

%and set the recovery matrix $\Gamma$ with uniform recovery rate $\gamma=\frac{1}{3}$. We classify the transmission rate ($b_{ij}$ for any node $i$ and $j$) into three categories, i.e., low/medium/high travel flow. Specifically, if $b_{ij}\in(0,0.01]$ 

%% file: Conclusions.tex
\vspace{-1.5ex}
\section{Conclusions} 
\vspace{-1ex}
\label{sec:conclusions}
This paper presents an input perturbation mechanism that provides differential privacy to graph weights when computing the basic reproduction number of an epidemic spreading network. 
The proposed mechanism uses bounded noise and the corresponding privacy-accuracy tradeoffs are quantified. We also develop a concentration bound to evaluate privacy-accuracy tradeoffs in terms of the remaining susceptible population within a community when the proposed mechanism is applied to a networked SIS or SIR model. Future works include applications of the proposed privacy mechanism in the control of epidemic spreading.

%% file: Appendix_new.tex
\vspace{-1ex}
\appendix
\vspace{-1ex}
We first state results that will be used later in proofs. 
\begin{definition}[Sub-Gaussian random variable\cite{buldygin1980sub}]\label{def:subG}
    A random variable $X\in\mathbb{R}$ is sub-Gaussian with variance proxy $\sigma_{SG}$, written as $X\sim\text{subG}(\sigma_{SG}^2)$, if $\mathbb{E}[X]=0$ and 
    %\begin{equation}
        $\mathbb{E}[\exp(\eta X)] \leq \exp\left(\frac{\eta^2\sigma_{SG}^2}{2}\right)$,
    %\end{equation}
    where $\mathbb{E}[\exp(\eta X)]$ is the moment generating function of random variable $X$. \hfill $\lozenge$
\end{definition}

%The tail bound on the sub-Gaussian random variables can be expressed in terms of the next lemma.
%\begin{lemma}\cite[Equation (2.9)]{wainwright_2019}\label{lem:subG_property}
%    If $X\sim \text{subG}(\sigma_{SG}^2)$, then
%    \begin{equation}
%        \mathbb{P}[X>t] \leq \exp\left(-\frac{t^2}{2\sigma_{SG}^2}\right), \forall t\geq 0. \ \ \ \ \ \ \  \ \ \ \ \ \ \ \ \ \ \ \ \ \ \ \ \ \ \ \ \ \ \ \ \ \hfill \blacksquare
%    \end{equation}
    %for any $t>0$.
%\end{lemma}

A truncated Gaussian random variable can be expressed as a sub-Gaussian random variable.

%\baike{[Baike: Are $\phi()$ and $\Phi()$ defined?]}\bo{Yes}
\begin{lemma}\label{lem:trunc_gauss_is_sub_gauss}
    If $Y\sim\text{TrunG}(\mu,\sigma,l,u)$, then we have that ${Y-\mathbb{E}[Y]\sim \text{subG}(\sigma^2)}$.
\end{lemma}

\noindent\emph{Proof of Lemma~\ref{lem:trunc_gauss_is_sub_gauss}:}
The proof of Lemma~\ref{lem:trunc_gauss_is_sub_gauss} will use the following lemma. 

\begin{lemma}\cite[Chapter 3]{burkardt2014truncated}\label{lem:truncated_gaussian_moments}
    If $Y\sim\text{TrunG}(\mu,\sigma,l,u) $, then 
    \begin{align}
        &\mathbb{E}[Y] = \mu + \sigma\cdot\frac{\varphi\left(\alpha\right)-\varphi\left(\beta\right)}{\Phi\left(\beta\right)-\Phi\left(\alpha\right)},\label{eq:trunc_gauss_mean}\\
        &\text{Var}[Y] = \sigma^2\left[1-\frac{\beta\varphi(\beta)-\alpha\varphi(\alpha)}{\Phi\left(\beta\right)-\Phi\left(\alpha\right)} - \left(\frac{\varphi\left(\alpha\right)-\varphi\left(\beta\right)}{\Phi\left(\beta\right)-\Phi\left(\alpha\right)}\right)^2\right],\label{eq:trunc_gauss_var}\\
        &\mathbb{E}[\exp(\eta X)]= \exp\left(\mu \eta+\frac{\sigma^2\eta^2}{2}\right) \\
        &\qquad\qquad\qquad\quad\qquad\cdot\left[\frac{\Phi\left(\beta-\sigma \eta\right)-\Phi\left(\alpha-\sigma \eta\right)}{\Phi\left(\beta\right)-\Phi\left(\alpha\right)}\right]\label{eq:trunc_gauss_moment_gen},
    \end{align}
    where $\alpha=\frac{l-\mu}{\sigma}$ and $\beta=\frac{u-\mu}{\sigma}$.\hfill $\blacksquare$
\end{lemma}

The proof then proceeds by extending the moment generating function of $Y$. By~\eqref{eq:trunc_gauss_moment_gen} we have
\vspace{-1ex}
\begin{equation}
    \mathbb{E}[\exp(\eta Y)] = \exp\left(\mu \eta+\frac{\sigma^2\eta^2}{2}+ f(\eta)\right),
\end{equation}
where $f(\eta)=\log\left(\frac{\Phi\left(\beta-\sigma \eta\right)-\Phi\left(\alpha-\sigma \eta\right)}{\Phi\left(\beta\right)-\Phi\left(\alpha\right)}\right)$. We now take the Taylor expansion of $f$ about $\eta=0$
to find 
\begin{equation}
    f(\eta) = f(0) + \left(\frac{\partial f(\eta)}{\partial \eta}\Bigr|_{\eta=0}\right)\eta+\frac{1}{2}\left(\frac{\partial^2 f(\eta)}{\partial \eta^2}\Bigr|_{\eta=\delta}\right)\delta^2,
\end{equation}
where $\delta\in(0,\eta)$. 
%\mh{Do we want this to be~$\delta \in [0, \infty)$?}
We note that $f(0)=0$ and the first derivative evaluated at $\eta=0$ is
\begin{align}
    \frac{\partial f(\eta)}{\partial \eta}\Bigr|_{\eta=0} 
    %&= \left(\sigma\cdot\frac{\varphi(\alpha-\sigma \eta)-\varphi(\beta-\sigma \eta)}{\Phi(\beta-\sigma \eta)-\Phi(\alpha-\sigma \eta)}\right)\Bigr|_{\eta=0}\\
    &= \sigma\cdot\frac{\varphi(\alpha)-\varphi(\beta)}{\Phi(\beta)-\Phi(\alpha)}.
\end{align}
For all $\eta>0$, we have $\frac{\partial^2 f(\eta)}{\partial \eta^2}\leq 0.$
%\begin{align}
    %\frac{\partial^2 f(\eta)}{\partial \eta^2}&=-\sigma^2\cdot\frac{(\beta-\sigma \eta)\varphi(\beta-\sigma \eta)-(\alpha-\sigma \eta)\varphi(\alpha-\sigma \eta)}{\Phi(\beta-\sigma \eta)-\Phi(\alpha-\sigma \eta)}\\
    %&-\sigma^2\cdot\left(\frac{\varphi(\alpha-\sigma \eta)-\varphi(\beta-\sigma \eta)}{\Phi(\beta-\sigma \eta)-\Phi(\alpha-\sigma \eta)}\right)^2\leq 0 .
    %\frac{\partial^2 f(\eta)}{\partial \eta^2}\leq 0.
    %\vspace{-2ex}
%\end{align}
Therefore, 
%\vspace{-2ex}
\begin{equation}
    f(\eta) \leq \sigma \eta\cdot\frac{\varphi(\alpha)-\varphi(\beta)}{\Phi(\beta)-\Phi(\alpha)}.\label{eq:taylor_series}
\end{equation}

We then conclude by extending the moment generating function of $Y-E[Y]$ and plugging in~\eqref{eq:trunc_gauss_mean} and~\eqref{eq:taylor_series}
to find 
%\begin{align}
    $\mathbb{E}[\exp(\eta(Y-\mathbb{E}[Y]))] = \exp(-\eta\mathbb{E}[Y])\mathbb{E}[\exp(\eta Y)] 
    \leq \exp\left(\frac{\sigma^2\eta^2}{2}\right)$.\hfill  $\blacksquare$
%\end{align}
\vspace{-1ex}
\subsection{Proof of Theorem~\ref{thm_upper_bound}}\label{apdx:proof_equilibrium_bound}
 Networked $SIS$ and $SIR$ models are special cases of the coupled system in~\cite[Eq.~(2)-(4)]{zhen2023steady}, as described in~\cite[Fig. 1]{zhen2023steady}. We leverage~\cite[Thm. 5]{zhen2023steady} for the coupled system in~\cite[Eq.~(2)-(4)]{zhen2023steady} to study the networked $SIS$ and $SIR$ models in~\eqref{eq:non_dyn_epi_s}-\eqref{eq:non_dyn_epi_x}. 
In order to leverage~\cite[Thm. 5]{zhen2023steady}, we investigate if the system~\eqref{eq:non_dyn_epi_s}
-\eqref{eq:non_dyn_epi_x} satisfies ~\cite[Assump. 3 and 4]{zhen2023steady}, and
the preconditions of the dynamics in~\cite[Eq.~(2)-(4)]{zhen2023steady}. 

We discuss networked $SIR$ models first.
Networked compartmental models such as $SIR$ models are groups of
polynomial ODEs over the compact set $[0,1]^n$. Hence,~\eqref{eq:SIR}
are locally Lipschitz in states $s(t)$, $x(t)$, $r(t)$, which satisfies the condition that the system dynamics in 
~\cite[Eq. (2)]{zhen2023steady} must be locally Lipschitz in their states.
Further, $I>0$ and $B>0$~in \eqref{eq:non_dyn_epi_x} guarantee that the nonlinear dynamics of $s(t)$ in~\eqref{eq:non_dyn_epi_s} satisfy the definition of~\cite[Eq.~(2)]{zhen2023steady}, where
$M_1 > 0$ and~$M_2 > 0$ in~\cite[Eq. (2)]{zhen2023steady}. 
For networked $SIR$ models, the states of each community obey $s_i(t), x_i(t)\in [0,1]$ for all~$i \in [n]$.  Therefore, the LTI system with external input $u(t)$ that captures the infected states $x(t)$ in~\eqref{eq:non_dyn_epi_x} is internally positive. In addition, $-\Gamma$ is a Hurwitz and Metzler matrix, since $-\Gamma$ is a negative definite diagonal matrix, where the $i^{th}$ diagonal entry is 
the recovery rate $\gamma_i$ of community $i$. By comparing~\cite[Eq. (3)]{zhen2023steady}  and~\eqref{eq:non_dyn_epi_x}, $I>0$ is a nonzero, nonnegative matrix. 
Therefore, the preconditions in~\cite[Eq.~(2)]{zhen2023steady}, namely that $B>0$ and $C>0$, hold here. 

Next, we check if~\eqref{eq:non_dyn_epi_s}-\eqref{eq:non_dyn_epi_x} satisfy~\cite[Assump. 3 and 4]{zhen2023steady}. 
The ~\cite[Assump. 3]{zhen2023steady} considers $(x^*,u^*)$  
such that $\dot x = 0$ in~\eqref{eq:non_dyn_epi_x}, and such that 
for sufficiently small $\nu>0$, it holds that $u\neq u^*$ and $\|u-u^*\|<\nu$ imply $\dot x \neq 0$ at $x^*$ and $u$.
%\mh{Should this say ``$\dot{x} \neq 0$''?}
%\mh{Don't use~$\epsilon$ here since that's a privacy parameter. Use another symbol.}

%Assumptions~\ref{assum:equi} and~\ref{assum:inv}. 
Consider~\eqref{eq:non_dyn_epi_x}. When $\dot x = 0$, we have that $-\Gamma x^* + Iu^*=0$. For any $u$ such that $\|u-u^*\|<\nu$, where $\nu>0$ can be sufficiently small, we have that $-\Gamma x^* - Iu =0- \Gamma x^* - Iu = I (u^*-u)$.
Note that we use the fact that~\eqref{eq:non_dyn_epi_x} is a linear system. Since $I\neq0$ and $\|u-u^*\|<\nu$, we have $\dot x = -\Gamma x^* + Iu\neq 0$. Thus, 
%we show that if  $-\Gamma x^* + Iu^*=0$, for any $u$ such that $\|u-u^*\|<\epsilon$, where $\epsilon>0$ can be sufficiently small, $-\Gamma x^* + Iu\neq 0$. We prove~
\cite[Assump. 3]{zhen2023steady} holds for~\eqref{eq:non_dyn_epi_x}.
%Proposition~\ref{prop:coupled} relies on. 
% %Consider networked $SIR$ models.
% Networked $SIR$ models  have infinite many disease-free equilibria where $i^*=0$, and $s^*=0$ $\forall i\in[n]$ or $s^*\neq 0$ $\forall i\in[n]$. For all the disease-free equilibria, 
% we have that 
% $\dot s_i(t) = -s^*_i(t)\sum_{i=1}^n \beta_{ij} x^*_{i}(t)=0$ $\forall i\in[n]$. 
% From this group of ODEs, at any equilibrium where $s^*_i\neq0$ $\forall i\in[n]$, we must have that $x^*_i=0$ $\forall i\in[n]$. %Consider networked $SIS$ models.  Networked $SIS$ models can have the disease-free equilibrium where $s^*_i=1$ and $x^*_i=0$ $\forall i\in[n]$ and the endemic equilibrium where $s^*_i\neq0$ and $x^*_i\neq0$ $\forall i\in[n]$.
% %We have that $\dot s_i(t) = -s_i(t)\sum_{i=1}^n \beta_{ij} x_i(t)+\gamma x_i(t)$. Recall that networked $SIS$ model can have either
% %At an equilibrium
% %where $s^*_i\neq0$ $\forall i\in[n]$, we cannot ensure $x^*_i=0$ $\forall i\in[n]$ since networked $SIS$ can have both disease-free equilibrium ($x^*_i=0$ $\forall i\in[n]$) and endemic equilibrium ($x^*_i\neq0$ $\forall i\in[n]$). 
% Hence, to establish Assumption~\ref{assum:equi} %in Proposition~\ref{prop:coupled}, 
% we consider disease-free equilibria networked $SIR$ models, where $s^*\neq0$, i.e., none of the community will become fully infected. 
%all possible equilibria for networked $SIR$ models and disease-free equilibria for networked $SIS$ models.

%\mh{This definition of~$\mathcal{I}$ is hard to follow. Can you re-phrase it?}
As for~\cite[Assump. 4]{zhen2023steady},
we need to investigate the set of the initial conditions of interest $\mathcal{I}$. If $(s(0),x(0))\in \mathcal{I}$, $\lim_{t\rightarrow\infty} s(t)=s^*$ exists. Further, $\mathcal{I}$ must satisfy the condition that if $(s(0),x(0))\in \mathcal{I}$, we can always find 
a $\tau>0$ such that for any $x>0$, $(s^*,\tau x)\in\mathcal{I}$.

% where for $f(s,x)$ in~\eqref{eq:non_dyn_epi_s}, 
%   if $(s,x)\in \mathcal{I}$, where $\mathcal{I}$ is the set of initial conditions of interest that guarantees the existence of $\lim_{t\rightarrow\infty} s(t)=s^*$. 
%   Then for all $x>0$, there exist a $\tau>0$ such that
%   $(s^*,\tau x)\in\mathcal{I}$. 
  
Without considering the dynamics of $r(t)$, for networked $SIR$ models, $\lim_{t\rightarrow\infty} s(t) = s^*$ exists if $(s(0),x(0))\in [0,1]^{2n}$~\cite{pare2020modeling}. 
Therefore, we assume $\mathcal{I} = [0,1]^{2n}$.
Then we show that if $\mathcal{I} = [0,1]^{2n}$ satisfies~\cite[Assump. 4]{zhen2023steady}.
We consider two sets of initial conditions $\mathcal{I}_1, \mathcal{I}_2\subset [0,1]^{2n}=\mathcal{I}$, where $\mathcal{I}_1\cap \mathcal{I}_2= \emptyset$ and $\mathcal{I}_1\cup \mathcal{I}_2= \mathcal{I}$. We define $\mathcal{I}_1 = \{ (s(0)=1,x(0)=0)\}$ that describes a trivial initial condition where no epidemic occurs, i.e., $\lim_{t\rightarrow\infty} s(t) = s^*= 1$ and $x^* = 0$. Under this condition, we cannot find a $\tau>0$ such that for any $x>0$, $(1,\tau x) \in \mathcal{I}_1$.
%Therefore, based on~\eqref{eq:couple}, we have $v=x=0$ that violates the condition that $v>0$ in Assumption~\ref{assum: Non}. 
Hence, although $\mathcal{I}_1\subset \mathcal{I}$, it does not satisfy ~\cite[Assump. 4]{zhen2023steady}.  Consider the set $\mathcal{I}_2 =[0,1]^{2n}\backslash \mathcal{I}_1 $. The set $\mathcal{I}_2$ describes the  initial conditions where at least one community has nonzero infections. For networked $SIR$ models, an initial condition in $\mathcal{I}_2$ guarantees $s^*<1$~\cite{mei2017dynamics,pare2020modeling}. For any $x>0$, we can always find a $\tau>0$ such that $s^* +\tau x < 1$, since $s^*<1$. 
%In addition, $r(0)=1-s^*-\tau x \in (0,1)$.
Hence, we can find a valid initial condition $(s^*,\tau x)\in\mathcal{I}_2$. We prove that
the non-trivial initial conditions in $\mathcal{I}_2$ for networked $SIR$ models always satisfy~\cite[Assump. 4]{zhen2023steady}.
Note that when we study networked $SIR$ models, we always consider initial conditions from $\mathcal{I}_2$.
%$v = y= x$ in~\eqref{eq:couple}, $v_i$
%for initial conditions in $I_2$, we can always find a $\tau>0$ such that $(s,\tau v)$ for all $v>0$.  

%For networked $SIR$ models, we always consider the initial conditions to satisfy
%$s_i(0)\in[0,1]$ $x_i(0)\in[0,1]$, and $(r_i(0)= 1-s_i(0)-x_i(0))\in[0,1]$ $\forall i \in[n]$. %Similar for networked $SIR$ models with states $r_i(0)$ $\forall i \in[n]$.

%If the initial conditions $s(0),x(0)$ of networked epidemic models captured by $\eqref{eq:non_dyn}$ are in the set $\mathcal{I}_2=[0,1]^{3n}/\{s=1\}$, then, we can always find a $\tau>0$ such that $(s,\tau v)$ for all $v>0$. Consider community $i\in[n]$, we have that $s_i(0)+x_i(0)+r_i(0)=1$. If $s_i(0) = 1$, $x_i(0)=0$ and $r_i(0)=0$. 

%($[0,1]^{3n}$ for networked $SIR$ models), 
%For networked compartmental models in this work, 
%if the initial conditions are in $[0,1]$, 
%the state trajectory will be in the same invariant set~\cite{pare2020modeling}. Therefore, Networked $SIR$ models meet Assumption~\ref{assum:inv}. %in~\eqref{eq:non_dyn_epi}, 

Recall that we represent networked $SIR$ models as the coupled systems in~\eqref{eq:non_dyn_epi_s}-\eqref{eq:non_dyn_epi_x}.
After proving that the networked $SIR$ dynamics captured by~\eqref{eq:non_dyn_epi_s}-\eqref{eq:non_dyn_epi_x} meet~\cite[Assump. 3 and 4]{zhen2023steady} and the predictions,
%Assumptions~\ref{assum: Non}-\ref{assum:inv},
%for~\eqref{eq:non_dyn_epi}, 
we can implement 
~\cite[Them. 1]{zhen2023steady}
%Proposition~\ref{prop:coupled} 
to study disease-free equilibria of the networked $SIR$ models.
Since the transmission network is undirected and connected, 
we can treat the underlying graph as being directed and strongly connected
by replacing each undirected edge with two directed edges
in the obvious way. 
Further,
the transmission matrix $B$ is a symmetric irreducible matrix. 
%we consider the spreading network to be strongly connected.
The healing matrix $\Gamma$ is a positive definite diagonal matrix, resulting in $W= B\Gamma^{-1}$ being a Metzler and irreducible matrix.
%through by~\eqref{eq:non_dyn_epi_s}-\eqref{eq:non_dyn_epi}. %where $s_i^*\neq 0$ $\forall i\in[n]$.
Through~\cite[Them. 1]{zhen2023steady}, at a disease-free equilibrium %excluding $s_i^*=0$ $\forall i\in[n]$ 
of the networked $SIR$ model, there must exist a community within the network whose susceptible proportion is given by 
\vspace{-4ex}
\begin{align}
\label{eq:u_bound}
    s^*_i&\leq\frac{1}{\rho(-M_2CA^{-1}FM_1)}
    %=\frac{1}{\rho(-B\Gamma^{-1})}
    =\frac{1}{\rho(-\Gamma^{-1}B)}
    =\frac{1}{R_0}.
    \vspace{-4ex}
\end{align}
%Further, under the condition that the transmission network is undirected, we obtain that the graph is strongly connected. Hence,
%the transmission matrix $B$ is a symmetric irreducible matrix. 
%We further derive the upper bound from~\eqref{eq:u_bound} to obtain 
% \begin{align}
% \label{eq:u_bound_final}
%     s^*_i&<\frac{1}{\rho(-B\Gamma^{-1})}=\frac{1}{\rho(-B\Gamma^{-1})^\top}\\
%     &=\frac{1}{\rho(-\Gamma^{-1}B^\top)}
%     =\frac{1}{R_0}.
% \end{align}
This completes the proof for networked $SIR$ models. 

For networked $SIS$ models, we show a straightforward conclusion.
%At the disease-free equilibrium, Assumption~\eqref{assum:inv} does not establish 
We know that $s^*=1$ and $x^*=0$ is the unique disease-free equilibrium for networked $SIS$ models~\cite{pare2020modeling,mei2017dynamics}. Further, networked $SIS$ models will reach the disease-free equilibrium if and only if $R_0\leq1$~\cite{pare2020modeling}. Hence, we have that for networked $SIS$ models at their disease-free equilibrium, $s^*_i=1\leq \frac{1}{R_0}$ for all $i\in [n]$. \hfill $\blacksquare$
\vspace{-4ex}
\subsection{Proof of Theorem~\ref{thm:expectation_bound_input_perturb}}\label{apdx:proof_expectation_bound_input_perturb}
Let $\Tilde{W}=W+P$, where the matrix $P$ encodes the differences between $\Tilde{W}$ and $W$. Then we have the equalities
%\begin{align}
    $\Tilde{R}_0-R_0 = \rho(\Tilde{W})-\rho(W)= \rho(W+P) - \rho(W)$.
%\end{align}
Since both $W$ and $P$ are symmetric, we apply~\cite[Theorem 8.4.11]{Bernstein2005MatrixMT} to split $\rho(W+P)$ and then apply Weyl's inequality~\cite[Fact 9.12.4]{Bernstein2005MatrixMT} to find 
\begin{align}   
    |\rho(W+P) - \rho(W)|\leq ||P||_2\leq||P||_F\label{ineq:eigenvalue_Frobenius},
\end{align}
where $||\cdot||_2$ is the $L_2$-norm of a matrix. Now we take the expectation on both sides of \eqref{ineq:eigenvalue_Frobenius} and apply Jensen's inequality to find 
\begin{align}
    \mathbb{E}[|\Tilde{R}_0 &- R_0|]
    \leq \mathbb{E}[||P||_F]\\
    &= \mathbb{E}\left[\sqrt{\sum_{i=1}^n\sum_{j=1}^n(p_{ij})^2}\right]
    \leq \sqrt{\mathbb{E}\left[\sum_{i=1}^n\sum_{j=1}^n(p_{ij})^2\right]}\\
    &= \sqrt{\sum_{i=1}^n\sum_{j=i+1}^n2\mathbb{E}\left[(p_{ij})^2 \right] + \sum_{i=1}^n \mathbb{E}\left[(p_{ii})^2\right]},\label{eq:split_expectation}
    %&= \sqrt{\sum_{i=1}^n\sum_{j=1}^n(\mathbb{E}\left[p_{ij}\right])^2+\text{Var}[p_{ij}]},
\end{align}
where $p_{ij}$ is the $i^{th},j^{th}$ entry of $P$. Equation~\eqref{eq:split_expectation} holds because each $p_{ij}$ above the main diagonal is generated independently and the matrix $P$ is symmetric. We note that $p_{ij}\sim\text{TrunG}(0,\sigma,\underline{w}_{ij}-w_{ij},\Bar{w}_{ij}-w_{ij})$. We apply the equality $\mathbb{E}[X^2]=(\mathbb{E}[X])^2+\text{Var}[X]$ for any random variable $X$ and $\mathbb{E}[p_{ij}]=\text{Var}[p_{ij}]=0$ if $p_{ij}=0$ to have
\begin{multline}
\vspace{-4ex}
\quad(\mathbb{E}\left[p_{ij}\right])^2+\text{Var}[p_{ij}] \\
    =\sigma^2\left(1-\frac{\beta_{ij}\varphi(\beta_{ij})-\alpha_{ij}\varphi(\alpha_{ij})}{\Phi(\beta_{ij})-\Phi(\alpha_{ij})}\right)\cdot\mathbf{1}_{\mathbb{R}_{> 0}}(w_{ij}),\label{eq:variance_relation}
\end{multline}
where $\mathbb{E}[p_{ij}]$ and $\text{Var}[p_{ij}]$ can be found by Lemma~\ref{lem:truncated_gaussian_moments}. Additionally, by~\eqref{ineq:eigenvalue_Frobenius} and $\text{Var}[X]\leq \mathbb{E}[X^2]$ for any $X$, we have
\begin{align}
    \text{Var}[|\Tilde{R}_0-R_0|]&\leq \mathbb{E}[||P||_F^2]\leq \mathbb{E}\left[\sum_{i=1}^n\sum_{j=1}^n(p_{ij})^2\right].
\end{align}
We conclude by applying~\eqref{eq:split_expectation} and~\eqref{eq:variance_relation}. \hfill $\blacksquare$

\vspace{-1ex}
\subsection{Proof of Theorem~\ref{thm:level_of_penetration_acc_bound_input_perturb}}\label{apdx:proof_level_of_penetration_acc_bound_input_perturb}

We introduce the following lemma for Theorem~\ref{thm:level_of_penetration_acc_bound_input_perturb}.
 \vspace{-1ex}
\begin{lemma}\cite[Corollary 4.4.8]{vershynin_2018}\label{lem:concentration_bound_random_matrix}
    Let $H$ be an $n\times n$ random symmetric matrix whose entries $H_{ij}$ on and above the main diagonal are independent, mean zero, sub-Gaussian random variables with proxy $\sigma^2$. Then, for any $v>0$ we have
    %\begin{equation}
        $\mathbb{P}\big[||H||_2\geq t\big] \leq 4\exp\left(-\frac{v^2}{2\sigma^2}\right)$,
    %\end{equation}
    where $t = \sqrt{2\sigma^2(4.4n+v^2)}$.\hfill $\blacksquare$
\end{lemma}

For a private undirected graph $\Tilde{G}=(V,E,\Tilde{W})$, we write %partition the private weight matrix $\Tilde{W}$ with
%\begin{equation}\label{eq:tilde_w_partition}
    $\Tilde{W}=W + M + A$,
%\end{equation}
%\vspace{-1ex}
where $M=[m_{ij}]^{n\times n}$ is a deterministic symmetric matrix with $m_{ij}=\mathbb{E}[\Tilde{w}_{ij}-w_{ij}]$. The matrix $A=[a_{ij}]^{n\times n}$ captures all the random parts with $a_{ij}\sim\text{TrunG}(-m_{ij},\sigma,\underline{w}_{ij}-w_{ij}-m_{ij},\Bar{w}_{ij}-w_{ij}-m_{ij})$. Note that $\mathbb{E}[a_{ij}]=0$ for every $i,j \in [n]$. Applying Weyl's inequality~\cite[Fact 9.12.4]{Bernstein2005MatrixMT} to obtain the inequality
%\begin{align}
    $|\rho(\Tilde{W}) -\rho(W)|\leq ||\Tilde{W}-W||_2\leq ||M||_F + ||A||_2$.
%\end{align}
Based on~\eqref{eq:trunc_gauss_mean}, 
%\mh{Where is~$T$ used here?}
\begin{align}
    ||M||_F &= \sqrt{\sum_{i=1}^n\sum_{j=1}^n m_{ij}^2}
    = \sqrt{\sum_{i=1}^n\sum_{j=1}^n (\mathbb{E}[\Tilde{w}_{ij}-w_{ij}])^2},
    %&= \sqrt{\sum_{i=1}^n\sum_{j=1}^n\left(\sigma\cdot\frac{\varphi\left(\alpha_{ij}\right)-\varphi\left(\beta_{ij}\right)}{\Phi\left(\beta_{ij}\right)-\Phi\left(\alpha_{ij}\right)}\right)^2\cdot\mathbf{1}_{T}(w_{ij})}.
\end{align}
where $(\mathbb{E}[\Tilde{w}_{ij}-w_{ij}])^2$ can be found using~\eqref{eq:trunc_gauss_mean}. Since the matrix~$A$ is symmetric, by Lemma~\ref{lem:concentration_bound_random_matrix} we have
 \vspace{-1ex}
\begin{multline}
\mathbb{P}\Big[|\rho(\Tilde{W}) - \rho(W)|> t+||M||_F\Big] \\
    \quad\leq \mathbb{P}\Big[||A||_2+||M||_F>t+||M||_F\Big] 
    %&= \mathbb{P}\Big[ ||A||_2 > t \Big]
    \leq 4\exp\left(-\frac{v^2}{2\sigma^2}\right).
\end{multline}
 \vspace{-1ex}
We conclude the proof by rearranging the terms above. \hfill $\blacksquare$

% \subsection{Proof of Theorem~\ref{thm:level_of_penetration_acc_bound_output_perturb}}\label{apdx:proof_level_of_penetration_acc_bound_output_perturb}
% For $\Tilde{R}_0$ generated by Mechanism~\ref{mech:bounded_laplace}, we have
% \begin{align}
%     &\quad\mathbb{P}\left[\left|\frac{1}{\Tilde{R}_0}-\frac{1}{R_0}\right|<t\right] \\
%     &= \mathbb{P}\left[-t<\left(\frac{1}{\Tilde{R}_0}-\frac{1}{R_0}\right)<t\right]\\
%     &=\mathbb{P}\left[-t+\frac{1}{R_0}<\frac{1}{\Tilde{R}_0}<t+\frac{1}{R_0}\right]\\
%     &= \mathbb{P}\left[\frac{1}{t+\frac{1}{R_0}}<\Tilde{R}_0<\frac{1}{-t+\frac{1}{R_0}}\right].
% \end{align}

% We can conclude the proof by further simplifying the equality above.\hfill$\square$

%% file: root.bbl
\begin{thebibliography}{10}
\providecommand{\url}[1]{#1}
\csname url@samestyle\endcsname
\providecommand{\newblock}{\relax}
\providecommand{\bibinfo}[2]{#2}
\providecommand{\BIBentrySTDinterwordspacing}{\spaceskip=0pt\relax}
\providecommand{\BIBentryALTinterwordstretchfactor}{4}
\providecommand{\BIBentryALTinterwordspacing}{\spaceskip=\fontdimen2\font plus
\BIBentryALTinterwordstretchfactor\fontdimen3\font minus
  \fontdimen4\font\relax}
\providecommand{\BIBforeignlanguage}[2]{{%
\expandafter\ifx\csname l@#1\endcsname\relax
\typeout{** WARNING: IEEEtran.bst: No hyphenation pattern has been}%
\typeout{** loaded for the language `#1'. Using the pattern for}%
\typeout{** the default language instead.}%
\else
\language=\csname l@#1\endcsname
\fi
#2}}
\providecommand{\BIBdecl}{\relax}
\BIBdecl

\bibitem{brauer2008compartmental}
F.~Brauer, ``Compartmental models in epidemiology,'' \emph{Math. Epi.}, pp.
  19--79, 2008.

\bibitem{pare2020modeling}
P.~E. Par{\'e}, C.~L. Beck, and T.~Ba{\c{s}}ar, ``Modeling, estimation, and
  analysis of epidemics over networks: An overview,'' \emph{Annual Reviews in
  Control}, vol.~50, pp. 345--360, 2020.

\bibitem{she2021peak}
B.~She, H.~C. Leung, S.~Sundaram, and P.~E. Par{\'e}, ``Peak infection time for
  a networked {SIR} epidemic with opinion dynamics,'' in \emph{In Proc. 60th
  IEEE Conf. on Dec. and Contr.}\hskip 1em plus 0.5em minus 0.4em\relax IEEE,
  2021, pp. 2104--2109.

\bibitem{mei2017dynamics}
W.~Mei, S.~Mohagheghi, S.~Zampieri, and F.~Bullo, ``On the dynamics of
  deterministic epidemic propagation over networks,'' \emph{Ann. Rev. in
  Cont.}, vol.~44, pp. 116--128, 2017.

\bibitem{brockmann2013hidden}
D.~Brockmann and D.~Helbing, ``The hidden geometry of complex, network-driven
  contagion phenomena,'' \emph{Science}, vol. 342, no. 6164, pp. 1337--1342,
  2013.

\bibitem{bengio2020need}
Y.~Bengio, R.~Janda, Y.~W. Yu, D.~Ippolito, M.~Jarvie, D.~Pilat, B.~Struck,
  S.~Krastev, and A.~Sharma, ``The need for privacy with public digital contact
  tracing during the {COVID-19} pandemic,'' \emph{The Lancet Digital Health},
  vol.~2, no.~7, pp. e342--e344, 2020.

\bibitem{delamater2019complexity}
P.~L. Delamater, E.~J. Street, T.~F. Leslie, Y.~T. Yang, and K.~H. Jacobsen,
  ``Complexity of the basic reproduction number (${R_0}$),'' \emph{Emerging
  Infectious Diseases}, vol.~25, no.~1, p.~1, 2019.

\bibitem{aronson2020will}
J.~K. Aronson, J.~Brassey, and K.~R. Mahtani, ``When will it be over?”: An
  introduction to viral reproduction numbers, {$R_0$ and $R_e$},'' \emph{The
  Centre for Evidence-Based Medicine}, vol.~14, 2020.

\bibitem{imola2021locally}
J.~Imola, T.~Murakami, and K.~Chaudhuri, ``Locally differentially private
  analysis of graph statistics,'' in \emph{Proc. 30th USENIX security symposium
  (USENIX Security 21)}, 2021, pp. 983--1000.

\bibitem{Karwa2014Private}
V.~Karwa, S.~Raskhodnikova, A.~Smith, and G.~Yaroslavtsev, ``Private analysis
  of graph structure,'' \emph{ACM Trans. Database Syst.}, vol.~39, no.~3, oct
  2014.

\bibitem{Day2016Publishing}
W.-Y. Day, N.~Li, and M.~Lyu, ``Publishing graph degree distribution with node
  differential privacy,'' in \emph{Proc. 2016 Int. Conf. on Mana of Data}, ser.
  SIGMOD '16.\hskip 1em plus 0.5em minus 0.4em\relax New York, NY, USA:
  Association for Computing Machinery, 2016, p. 123–138.

\bibitem{ZHANG2021Differentially}
S.~Zhang, W.~Ni, and N.~Fu, ``Differentially private graph publishing with
  degree distribution preservation,'' \emph{Computers \& Security}, vol. 106,
  p. 102285, 2021.

\bibitem{chen2021edge}
B.~Chen, C.~Hawkins, K.~Yazdani, and M.~Hale, ``Edge differential privacy for
  algebraic connectivity of graphs,'' in \emph{Proc. 60th IEEE Conf. on Dec.
  and Cont. (CDC)}.\hskip 1em plus 0.5em minus 0.4em\relax IEEE, 2021, pp.
  2764--2769.

\bibitem{Bureau2023DP}
``Why the {Census Bureau} chose differential privacy,'' \emph{{U.S. Census
  Bureau}}, 2023.

\bibitem{Sealfon2016shortest}
A.~Sealfon, ``Shortest paths and distances with differential privacy,'' ser.
  PODS '16.\hskip 1em plus 0.5em minus 0.4em\relax New York, NY, USA:
  Association for Computing Machinery, 2016, p. 29–41.

\bibitem{pinot2018clustering}
R.~Pinot, A.~Morvan, F.~Yger, C.~Gouy-Pailler, and J.~Atif, ``{Graph-based
  Clustering under Differential Privacy},'' in \emph{{Proc. Conf. on Uncer. in
  Art. Int. (UAI 2018)}}, Monterey, California, United States, Aug. 2018, pp.
  329--338.

\bibitem{dwork2014algorithmic}
C.~Dwork, A.~Roth \emph{et~al.}, ``The algorithmic foundations of differential
  privacy,'' \emph{Foundations and Trends{\textregistered} in Theoretical
  Computer Science}, vol.~9, no. 3--4, pp. 211--407, 2014.

\bibitem{chen2022bounded}
B.~Chen and M.~Hale, ``The bounded {Gaussian} mechanism for differential
  privacy,'' \emph{arXiv preprint arXiv:2211.17230}, 2022.

\bibitem{hawkins2022node}
C.~Hawkins, B.~Chen, K.~Yazdani, and M.~Hale, ``Node and edge differential
  privacy for graph {Laplacian} spectra: Mechanisms and scaling laws,''
  \emph{arXiv preprint arXiv:2211.15366}, 2022.

\bibitem{Hay2009accurate}
M.~Hay, C.~Li, G.~Miklau, and D.~Jensen, ``Accurate estimation of the degree
  distribution of private networks,'' in \emph{Proc. 2009 Ninth IEEE Int. Conf.
  on Data Mining}, 2009, pp. 169--178.

\bibitem{blocki2013differentially}
J.~Blocki, A.~Blum, A.~Datta, and O.~Sheffet, ``Differentially private data
  analysis of social networks via restricted sensitivity,'' \emph{arXiv
  preprint arXiv:1208.4586}, 2013.

\bibitem{Kasiviswanathan2013Analyzing}
S.~P. Kasiviswanathan, K.~Nissim, S.~Raskhodnikova, and A.~Smith, ``Analyzing
  graphs with node differential privacy,'' in \emph{Theory of Cryptography},
  A.~Sahai, Ed.\hskip 1em plus 0.5em minus 0.4em\relax Berlin, Heidelberg:
  Springer Berlin Heidelberg, 2013, pp. 457--476.

\bibitem{Wang2013Differential}
Y.~Wang, X.~Wu, and L.~Wu, ``Differential privacy preserving spectral graph
  analysis,'' in \emph{Proc. Advan.s in Kno. Disc. and Data Min.}\hskip 1em
  plus 0.5em minus 0.4em\relax Berlin, Heidelberg: Springer Berlin Heidelberg,
  2013, pp. 329--340.

\bibitem{Chanyaswad2018MVG}
T.~Chanyaswad, A.~Dytso, H.~V. Poor, and P.~Mittal, ``Proc. {MVG} mechan.:
  Diff. priv. under matrix-valued query,'' in \emph{Proceedings of the 2018 ACM
  SIGSAC Conf. on Comp. and Commu. Secu.}, ser. CCS '18.\hskip 1em plus 0.5em
  minus 0.4em\relax New York, NY, USA: Assoc. for Comp. Mach., 2018, p.
  230–246.

\bibitem{zhen2023steady}
S.~Zhen~Khong and L.~Su, ``Steady-state analysis of networked epidemic
  models,'' \emph{arXiv e-prints}, pp. arXiv--2305, 2023.

\bibitem{jagielski2020auditing}
M.~Jagielski, J.~Ullman, and A.~Oprea, ``Auditing differentially private
  machine learning: How private is private {SGD}?'' \emph{arXiv preprint
  arXiv:2006.07709}, 2020.

\bibitem{nasr2021adversary}
M.~Nasr, S.~Song, A.~Thakurta, N.~Papernot, and N.~Carlini, ``Adversary
  instantiation: Lower bounds for differentially private machine learning,''
  \emph{arXiv preprint arXiv:2101.04535}, 2021.

\bibitem{song2019auditing}
C.~Song and V.~Shmatikov, ``Auditing data provenance in text-generation
  models,'' \emph{arXiv preprint arXiv:1811.00513}, 2019.

\bibitem{balle2022reconstructing}
B.~Balle, G.~Cherubin, and J.~Hayes, ``Reconstructing training data with
  informed adversaries,'' \emph{arXiv preprint arXiv:2201.04845}, 2022.

\bibitem{burkardt2014truncated}
J.~Burkardt, ``The truncated normal distribution,'' \emph{Department of
  Scientific Computing Website, Florida State University}, vol.~1, p.~35, 2014.

\bibitem{Minnesota_2023}
\BIBentryALTinterwordspacing
``Minnesota coronavirus cases and deaths,'' 2023. [Online]. Available:
  \url{https://doi.org/10.1145/237814.237880}
\BIBentrySTDinterwordspacing

\bibitem{brooks2023TNSEflows}
B.~A. Butler, R.~Stern, and P.~E. Par\'e, ``Analysis and applications of
  population flows in a networked {SEIRS} epidemic process,'' \emph{ArXiv},
  2023, arXiv:2309.11588 [math.OC].

\bibitem{buldygin1980sub}
V.~V. Buldygin and Y.~V. Kozachenko, ``Sub-gaussian random variables,''
  vol.~32, no.~6, pp. 483--489.

\bibitem{Bernstein2005MatrixMT}
D.~Bernstein, \emph{Matrix Mathematics: Theory, Facts, and Formulas (Second
  Edition)}, ser. Princeton reference.\hskip 1em plus 0.5em minus 0.4em\relax
  Princeton University Press, 2009.

\bibitem{vershynin_2018}
R.~Vershynin, \emph{High-Dimensional Probability: An Introduction with
  Applications in Data Science}, ser. Cambridge Series in Statistical and
  Probabilistic Mathematics.\hskip 1em plus 0.5em minus 0.4em\relax Cambridge
  University Press, 2018.

\end{thebibliography}
